%% file: main.tex
\documentclass[format=sigconf, screen, dvipsnames]{acmart}
\AtBeginDocument{%
  \providecommand\BibTeX{{%
    \normalfont B\kern-0.5em{\scshape i\kern-0.25em b}\kern-0.8em\TeX}}}

\copyrightyear{2025}
\acmYear{2025}
\setcopyright{acmlicensed}\acmConference[UIST '25]{The 38th Annual ACM Symposium on User Interface Software and Technology}{September 28-October 1, 2025}{Busan, Republic of Korea}
\acmBooktitle{The 38th Annual ACM Symposium on User Interface Software and Technology (UIST '25), September 28-October 1, 2025, Busan, Republic of Korea}
\acmDOI{10.1145/3746059.3747614}
\acmISBN{979-8-4007-2037-6/2025/09}

\usepackage{bbding}
\usepackage{fontawesome5}
\usepackage{pifont}
\usepackage{enumitem}
\usepackage{siunitx}
\usepackage{graphicx}
\usepackage{subcaption}
\usepackage{url}
\usepackage{multirow}

\DeclareSIUnit\sq{\ensuremath{\Box}}
\newcommand{\xmark}{\ding{55}}%
\begin{document}

\newcommand{\rev}[1]{#1}
% command to show revision highlighting
\definecolor{REVISIONRED}{HTML}{9e092f}
\definecolor{REVISIONGREEN}{HTML}{427a11}
\definecolor{REVISIONGOLD}{HTML}{806b06}
\newcommand{\showrevisions}[1]{
    \renewcommand{\rev}[1]{{\color{REVISIONRED}##1}}
}
% \showrevisions

\setlength{\fboxsep}{0pt} % No gap between color and border

\title{SparseEMG: Computational Design of Sparse EMG Layouts for Sensing Gestures}

\author{Anand Kumar}
\orcid{0009-0002-6387-0604}
\affiliation{%
  \institution{University of Calgary}
  \streetaddress{2500 University Dr NW}
  \city{Calgary}
  \country{Canada}
  \postcode{T2N 1N4}}
\email{anand.kumar1@ucalgary.ca}

\author{Antony Albert Raj Irudayaraj}
\orcid{0000-0003-2015-523X}
\affiliation{%
  \institution{University of Calgary}
  \streetaddress{2500 University Dr NW}
  \city{Calgary}
  \country{Canada}
  \postcode{T2N 1N4}}
\email{antony.irudayaraj@ucalgary.ca}

\author{Ishita Chandra}
\orcid{0009-0002-4693-328X}
\affiliation{%
  \institution{University of Calgary}
  \streetaddress{2500 University Dr NW}
  \city{Calgary}
  \country{Canada}
  \postcode{T2N 1N4}}
\email{ishita.chandra@ucalgary.ca}

\author{Adwait Sharma}
\orcid{0000-0001-5676-3136}
\affiliation{%
  \institution{University of Bath}
  \streetaddress{Claverton Down}
  \city{Bath}
  \country{United Kingdom}
  \postcode{BA2 7AY}}
\email{as5339@bath.ac.uk}

\author{Aditya Shekhar Nittala}
\orcid{0000-0002-3698-9733}
\affiliation{%
  \institution{University of Calgary}
  \streetaddress{2500 University Dr NW}
  \city{Calgary}
  \country{Canada}
  \postcode{T2N 1N4}}
\email{anittala@ucalgary.ca}


%\author{Anonymized for submission}

\renewcommand{\shortauthors}{Kumar, et al.}

\begin{abstract}
Gesture recognition with electromyography (EMG) is a complex problem influenced by gesture sets, electrode count and placement, and machine learning parameters (e.g., features, classifiers). Most existing toolkits focus on streamlining model development but overlook the impact of electrode selection on classification accuracy. In this work, we present the first data-driven analysis of how electrode selection and classifier choice affect both accuracy and sparsity. Through a systematic evaluation of 28 combinations (4 selection schemes, 7 classifiers), across six datasets, we identify an approach that minimizes electrode count without compromising accuracy. The results show that Permutation Importance (selection scheme) with Random Forest (classifier) reduces the number of electrodes by 53.5\%. Based on these findings, we introduce SparseEMG, a design tool that generates sparse electrode layouts based on user-selected gesture sets, electrode constraints, and ML parameters while also predicting classification performance. SparseEMG supports 50+ unique gestures and is validated in three real-world applications using different hardware setups. Results from our multi-dataset evaluation show that the layouts generated from the SparseEMG design tool are transferable across users with only minimal variation in gesture recognition performance.

\end{abstract}





\begin{teaserfigure}
  \includegraphics[width=\textwidth]{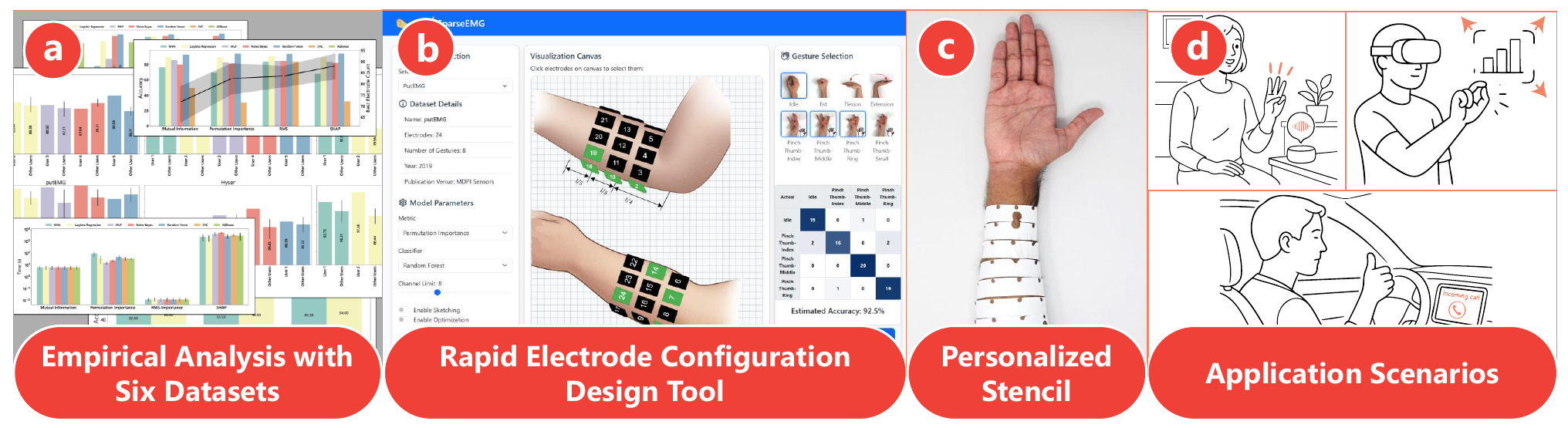}
  \caption{SparseEMG is a computational tool designed for rapid prototyping of sparse EMG electrode layouts for gesture recognition. Overview of our method (a) Empirical analysis across six datasets to identify optimal electrode selection strategies, (b) design tool that recommends sparse electrode configurations with high-level input, (c) and generates personalized stencils for user-specific electrode placement, and (d) evaluation across diverse real-world scenarios to demonstrate the tool's applicability.}
  \label{fig:teaser}
\end{teaserfigure}

\maketitle

\input{introduction}

\input{related_work}

\input{design_goals}
\input{method_v3}
\input{design_tool}
\input{applications}
\input{discussion}

\begin{acks}
This project received funding from the National Science and Engineering Research Council (NSERC) Canada (RGPIN - 2023-03608), NFRF (New Frontiers in Research Fund (NFRFE/00256-2022), A-MEDICO(Alberta Medical Devices Innovation Consortium) and Alberta Innovates Postdoctoral Fellowship.
\end{acks}

\bibliographystyle{ACM-Reference-Format}
\bibliography{sample-base}

\onecolumn
\appendix
\section{Appendix}

\begin{figure*}[h]
    \centering
    \includegraphics[width=1\linewidth]{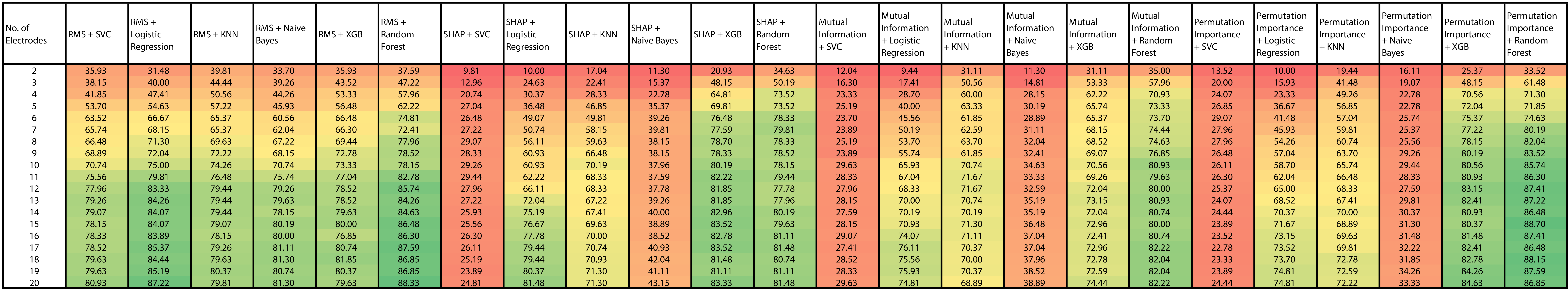}
    \caption{Raw data after setting a constraint of 20 electrodes for all the combinations}
    \label{fig:20_electrode_raw_data}
\end{figure*}








\end{document}

%% file: introduction.tex
\section{Introduction} 
Surface Electromyography (sEMG) is a technique to record muscular
activity by measuring the electrical field generated by the contractions of muscle fibres through electrodes placed on the skin. While extensively used in neuroscience, biomedical engineering, prosthetic control and rehabilitation studies, this technique is also used in HCI for gesture recognition. Seminal work by Saponas et al.~\cite{saponas_chi08, saponas_muscle_computer_chi10, saponas_uist09} and Constanza et al.~\cite{emg_intimate_interfaces_chi05,emg_intimate_interfaces_chi07} demonstrated the use of sEMG signals for gesture recognition, coining the term `muscle-computer interfaces.' Subsequently, plenty of follow-up work has studied gesture recognition with sEMG~\cite{user-independent-emg-recognition-mobilehci17,libEMG_ieee_access,embody_eics21}, and even consumer-grade EMG devices are being actively researched~\footnote{\url{https://tech.facebook.com/reality-labs/2021/3/inside-facebook-reality-labs-wrist-based-interaction-for-the-next-computing-platform/}}.

However, building EMG systems from scratch is complex and requires expertise in numerous domains, such as sensing technology, signal processing, machine learning, and interaction design. Sensing with EMG signals also requires placing the electrodes at precise locations for high-quality gesture recognition. This placement of electrodes becomes even more challenging when designing an optimal layout for recognizing a given set of gestures. 

Recent research has contributed open-source toolkits to facilitate the development of myoelectric control systems for non-domain experts. These toolkits abstract many challenges, including data processing, hardware interfacing, feature extraction/selection, classification, postprocessing, and evaluation~\cite{libEMG_ieee_access, embody_eics21}. While impressive, these toolkits focus on developing interactive ML systems for a given set of electrodes and gestures, i.e. they assume that a given set of electrodes are appropriately placed and then allow users to train a set of gestures. \textit{They do not focus on finding a sparse layout of electrodes for sensing a given set of gestures}.

Recent research in HCI has proposed the use of optimal sensor placement for sensing fine-grained microgestures~\cite{sparseIMU_tochi23}. There have been similar efforts which have shown that identifying the optimal location for sensor placement can boost the sensing accuracy~\cite{yizheng_uist19,ready_steady_touch_imwut20,backhand_uist15} or it can reduce the number of sensors required for detecting a given set of microgestures~\cite{sparseIMU_tochi23}. Similarly, Lin et al.~\cite{backhand_uist15} suggested that the minimum accuracy of 70.8\%  can be increased to 95.8\% for an identified optimal location. In a broader context, a large body of work has used compressed sensing and data-driven techniques to demonstrate that human body pose can be reconstructed by a significantly reduced number of sensors~\cite{reduced_marker_layouts_mig15,eckhoff2020sparse,deep_inertial_poser_siggraph18,real-time-physics-cvmp16}. However, all these efforts have been focused on sensors such as IMUs~\cite{sparseIMU_tochi23,deep_inertial_poser_siggraph18}, strain gauges~\cite{backhand_uist15}, or optical markers~\cite{reduced_marker_layouts_mig15} and do not generalize to sEMG sensing.

Biomedical and physical sciences research has used data-driven and signal-processing techniques to analyze sEMG signals for identifying the innervation zones~\cite{barbero2012atlas}. This was typically done by placing an extensive array of sEMG electrodes on the target muscle group and collecting the signals across the array. There have been similar efforts to identify the optimal placement of electrodes~\cite{jensen1996estimating,rainoldi2004method,optimized_electrode_embs2009,nittala2021computational}. However, these focus on rehabilitation, prosthesis control or for determining the major muscle groups that are responsible for specific muscular contractions (e.g.  Flexor Carpi Radialis flexes abducts the hand at the wrist~\cite{wheeles_fcr}). These techniques do not generalize to gestures because each gesture typically involves the contraction of multiple muscle groups, and no established mapping exists between a given gesture and a muscle group.

This paper addresses the following research questions, which, to the best of our knowledge, have not been explored in prior work which focused on EMG gesture recognition.
\begin{itemize}[leftmargin=*]

    \item How can we objectively quantify the involvement of specific electrodes for a given set of gestures?
    \item For a given set of constraints (number of channels, given gesture set, ML parameters), can we identify an electrode layout that can balance sparsity without compromising the recognition accuracy?
    \item How well do the solutions generated by our tool scale across widely available EMG hardware platforms, and for practical real-world scenarios?
\end{itemize}

\rev{
Exploring these research questions offers several benefits. Firstly, by determining the optimal number of electrodes needed for a specific set of gestures, we can simplify the hardware acquisition process and reduce costs. For instance, decreasing the number of electrodes from 16 to 6 can lower the price from several thousand dollars to just a few hundred by transitioning from professional-grade acquisition setups to low-cost Arduino-based EMG systems. Secondly, in an Arduino-based setup, having fewer channels allows for a higher frame rate, which can enhance performance. Finally, investigating these questions facilitates the creation of a computational tool that simplifies the complexities of machine learning and EMG. This empowers users to generate gesture-specific layouts with just a few clicks.
}
\begin{figure}
    \centering
    \includegraphics[width=1\linewidth]{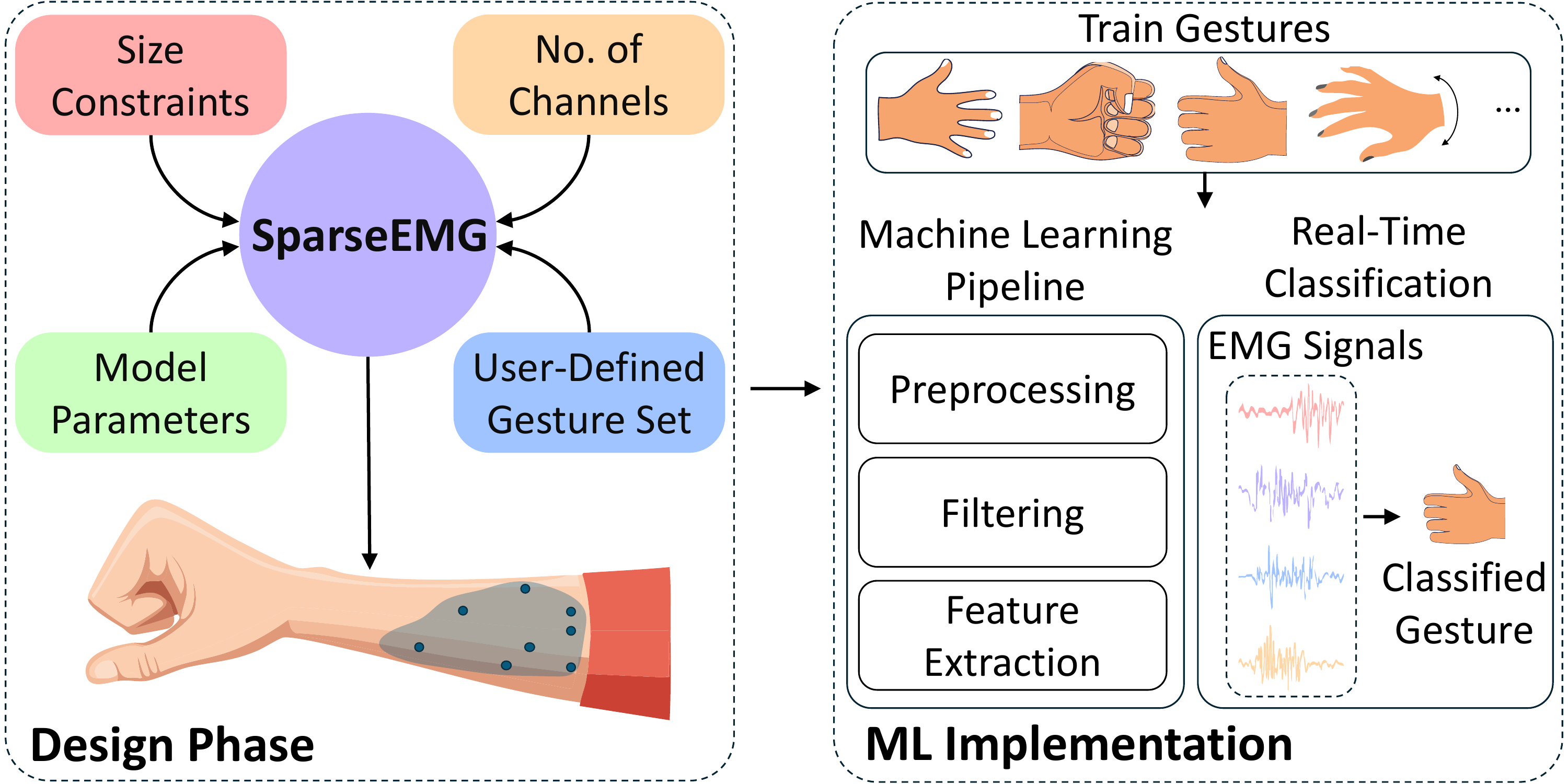}
    \caption{Existing EMG toolkits in HCI majorly focus on creating a machine learning pipeline for recognizing pre-defined gestures and easing this process~\cite{libEMG_ieee_access,embody_eics21} (which falls in the ML implementation phase of EMG gesture recognition). In contrast to these approaches, our design tool helps designers to rapidly select sparse EMG layouts for a desired set of gestures and other constraints specific to EMG and machine learning ( which usually occurs during the gesture and device design phase).  }
    \label{fig:enter-label}
\end{figure}

Overall, our paper makes the following contributions:

\begin{enumerate}
  \item \textbf{Series of Empirical Analysis}: We contribute a systematic data-driven analysis to understand how electrode selection and sparisty affect gesture recognition. Through a systematic evaluation of 28 combinations (4 selection
schemes, 7 classifiers) across six datasets to understand the best combination of electrode-selection metrics and ML classifiers for a given set of electrodes. 
 (Figure \ref{fig:teaser}(a))
  
  \item \textbf{Computational Design Tool}: We contribute a computational design tool that allows designers to identify an optimal set of electrodes for a given gesture set and a desired number of electrodes. The tool abstracts the designer from the lower-level technical aspects such as the anatomical relation between gestures and muscles, machine learning classifiers and sEMG signal characteristics. (Figure \ref{fig:teaser}(b))

  \item \textbf{Application Scenarios:} We contribute application scenarios from diverse and representative domains to illustrate  how designers and engineers can leverage the potential of our approach for concrete design tasks. Our application examples demonstrate that the computational approach is scalable to different hardware setups. (Figure \ref{fig:teaser}(d))

  % \item \textbf{Reduction in Electrode Count}: Identification of the minimum number of electrodes required for maximum accuracy, significantly reducing the computational and physical resource requirements of sEMG systems for gesture recognition.

  % \item \textbf{Development of a Tool for Electrode Optimization}: Development of a scalable tool capable of evaluating optimal electrode placement and number of electrode configurations for a given set of gestures.
\end{enumerate}

%% file: related_work.tex
%!TEX root = sample-manuscript.tex

\section{Related Work}
\begin{table*}
  \centering
  \fontsize{7}{11}\selectfont
  \begin{tabular}{p{0.10\textwidth} p{0.22\textwidth} p{0.14\textwidth} p{0.18\textwidth}  p{0.14\textwidth}  p{0.14\textwidth}}
    % \toprule
%     & & \multicolumn{2}{c}{\small{\textbf{Test Conditions}}} \\
%     \cmidrule(r){3-4}
    {\small  \textbf{Related Work}}
    & {\small\textbf{Context}}
    & {\small\textbf{Gesture Variability}}
    & {\small\textbf{Hardware Independent}}
    & {\small \textbf{Optimized Layouts}}
    & {\small \textbf{Dataset Scalability }
    } \\

    \toprule
    %& {\bf Single Substrate (on tattoo paper) } &  &  & \\
    %\midrule

      libEMG~\cite{libEMG_ieee_access} & Toolkit for EMG gesture recognition & \xmark & \checkmark & \xmark & \checkmark .\\

      EMBody\cite{embody_eics21} & Toolkit for EMG gesture recognition & \xmark & \xmark & \xmark & \xmark .\\

      SparseIMU~\cite{sparseIMU_tochi23} & Tool for Sparse IMU generation & \checkmark & \checkmark & \checkmark & \xmark .\\

      Nittala et al.~\cite{nittala2021computational} & Tool for optimized EMG layouts & \xmark & \checkmark & \checkmark & \xmark .\\

      Li et al.~\cite{li2024optimizing} & Optimizing the feature set and electrode configuration for EMG recognition & \xmark & \xmark & \checkmark & \xmark .\\

      BioPatRec~\cite{ortiz2013biopatrec} & Software for development of algorithms in prosthetic control & \xmark & \xmark & \xmark & \checkmark .\\

      Luis et al.~\cite{pelaez2022reducing} & Analyzes the impact of electrode configuration to reduce channel count & \xmark & \xmark & \checkmark & \xmark .\\

      Prakash et al.~\cite{prakash2025optimized} & Optimizing electrode configuration for wrist-worn hand gesture recognition. & \xmark & \xmark & \checkmark & \xmark .\\

      Fang et al.~\cite{electrode_configuration_ieee_tim_23} & Electrode configuration framework to improve the operational effectiveness & \checkmark & \xmark & \checkmark & \xmark .\\

      Zhao et al.~\cite{zhao2024adaptive} & Electrode optimization for prosthetics & \xmark & \xmark & \checkmark & \xmark .\\

      Chaplot et al.~\cite{electronics13153072} & Electrode optimization with five placement configurations & \xmark & \xmark & \checkmark & \xmark .\\ 

\midrule

      \textbf{This work} & Interactive exploration of gestures, electrode configurations and accuracy. & \checkmark & \checkmark & \checkmark & \checkmark .\\
     
    \bottomrule
  \end{tabular}
 \caption{ Comparison of related work across HCI, biomedical and prosthesis control that has contributed to the analysis of sEMG signals and the design of toolkits. }~\label{tab:related_work_table}
 
  %\descriuse ptionlabel{This table shows the materials, prices and their sourcing websites.}
  \vspace{-2em}
\end{table*}

Our work falls at the intersection of EMG sensing, computational design tools and toolkits for gesture sensing. 

\subsection{EMG Gesture Recognition in HCI}
The success of using electromyography (EMG) for prosthesis control led to its exploration by researchers in Human-Computer Interaction (HCI) in the early 2000s~\cite{saponas_chi08,saponas_muscle_computer_chi10,saponas_uist09,emg_intimate_interfaces_chi07,emg_intimate_interfaces_chi05}. Prior to this, surface EMG devices were expensive, tethered, and mainly used for medical purposes. The introduction of commercially available wearable devices, such as the Myo Armband in 2014~\cite{evaluating_myoband}, opened new possibilities for the broader research community to explore this technology for novel hands-free interactions. 

Since then, myoelectric control has been applied to various general-purpose applications, including piano augmentation~\cite{karolus2020hit}, drone control~\cite{emg_drone_control}, gaming~\cite{nittala2021computational,emg_gaming_chi17}, hands-free input~\cite{emg_same-side_interactions_chiea15}, and mixed-reality interactions~\cite{mobile_robot_control_emg}. At the same time, the HCI community has made significant contributions, applying their expertise to enhance prosthetics-related research through training tools~\cite{myo_prosthesis_CHI17,emg_vr_prosthesis}, approaches to alleviate phantom limb pain~\cite{phantom_limb_pain_emg_ahs22}, and improved prosthesis designs~\cite{deep_learning_prosthesis_imwut19}.
There is also increasing attention to EMG sensing from ML and computer vision communities ~\cite{neuroPose_www21,salter2024emg2pose,user-independent-emg-recognition-mobilehci17}. However, these works focus on user-independent models, pose estimation from a commercial EMG hardware band (e.g. MyoBand) and not on generating spare EMG layouts for a given set of gestures and other user-defined constraints. Despite the growing interest and adoption of this technology, both commercially~\cite{jaloza2021inside} and academically, the widespread use of myoelectric control for general applications remains limited~\cite{call_for_action_emg_chi23}. Recent work by Eddy et al.~\cite{call_for_action_emg_chi23} supports the notion that designing gesture recognition systems with EMG requires EMG-specific domain expertise, machine learning and interaction design. While there have been efforts in building toolkits to ease this process~\cite{libEMG_ieee_access,embody_eics21}, these focus on building ML pipelines and don't focus on the interaction design of gestures and generating a sparse EMG layout for a set of user-selected gestures. Our toolkit fills this gap in EMG toolkit research by abstracting the designers from EMG-specific domain expertise and instead focusing on gesture design for specific application scenarios.

\subsection{EMG in Biomedical and Physical Sciences}
EMG sensing has been extensively used in biomedical engineering and physical science research communities. Very early work in biomedical engineering used sEMG for understanding the mapping between limb movements and the muscles~\cite{barbero2012atlas}. More recently, using sEMG and machine learning for prosthesis control has received more attention in the biomedical research community, contributing datasets~\cite{ninapro_2012,putemg}, classifier-finetuning~\cite{niu2024optimizing,karnam2022emghandnet,deng2018discriminant} and open-source toolkits~\cite{libEMG_ieee_access,low-cost-hardware-toolkits-access2024,bioelectric_signal_bsn17}. BioPatRec is one of the earliest efforts in prosthesis control research that provides a modular, customizable platform that abstracts the practitioners from domain-specific knowledge in signal processing, feature selection, extraction, classification and real-time control~\cite{ortiz2013biopatrec}. Follow-up works have investigated the performance of various classifiers and contributed fine-tuned classifier topologies for improving the accuracy for prosthetic control~\cite{prahm2016combining,classifier_topologies_embc2013}. Similar efforts have contributed open-source datasets and benchmarking toolkits for analyzing the sEMG signals~\cite{open_access_systems_ieee_nsre_2021,sapsanis2014semg,khushaba2012toward}. 
There has been work on optimizing the number of sEMG channels for gesture recognition~\cite{prakash2025optimized,liu2022research}. However, these works focus on a specific dataset and for a pre-defined set of gestures and do not support the interactive exploration of variations in gesture set and user-defined electrode constraints. A more detailed review of feature extraction techniques, deep learning approaches for sEMG in the context of prosthetic control and clinical evaluation of neuromusculoskeletal systems can be found here~\cite{sid2024comprehensive}. While there is vast literature in biomedical and physical sciences research on using sEMG signals, they primarily focus on improving classification accuracy or incorporating deep-learning approaches or hardware/software toolkits for analyzing sEMG signals. To the best of our knowledge, we are yet to see a systematic analysis and design tool that enables interactive exploration of the various parameters that can influence sEMG signals.

\subsection{Computational Tools and Toolkits in HCI}
One solution to improve accessibility for complex areas of research is through the development of toolkits. Toolkits provide easy access to complex algorithms, enable fast prototyping of software and hardware interfaces, and/or enable creative exploration of design
spaces~\cite{evaluation_strategies_ledo_chi18}. Research in HCI has contributed toolkits~\cite{tacttongue_uist23,embody_eics21}, fabrication techniques~\cite{fabrication_toolkit,soft_sensors} and design tools for various applications, such as for electrical muscle stimulation~\cite{ems_toolkit_ah18,ems_twitch_ah13}, physiological sensing~\cite{physioskin_chi20}, actuation~\cite{anisma_tochi22,torque_capsules_uist24} and haptics~\cite{bAREfoot_uist20,tacttongue_uist23, vibrotactile_braille, tran_chi23, compressables_dis21}. Similarly, several design tools and elicitation studies have been presented in the HCI literature for the design of various gestures~\cite{danyluk_map_gestures,robust_microgestures_chi21,grasping_microgestures_chi19}. These include works that allow the designer to compare a gesture with a corpus of everyday activity data for false positive testing~\cite{magic_chi10, magic2_ieee_fg}. EventHurdle~\cite{eventhurdle_chi13}, M.Gesture~\cite{m_gesture_chi16}, and Mogeste~\cite{mogeste_indiahci16} enable users to compose custom gestures on mobile devices. Gesture Coder~\cite{gesture_coder_chi12} is a tool to help developers add multi-touch gestures by demonstrating them on a tablet’s touchscreen. While existing frameworks and platforms enable prototyping and debugging various classifiers and implementing custom machine learning pipelines~\cite{weka_toolkit,gestalt_uist10}, they are targeted at programmers and do not consider aspects of interaction design. On the other hand, recent technological advances have enabled novice users to train and classify custom ML models without the need for programming expertise~\cite{rapsai_chi23}. However, these majorly address image or audio classification problems. SparseIMU~\cite{sparseIMU_tochi23} contributes a design tool for generating sparse IMU layouts for gesture recognition. However, it does not scale to EMG sensing which is a completely different sensing mechanism. Secondly, the number of IMUs in SparseIMU is 17, while EMG arrays typically involve a much denser array of electrodes ranging from 10 to 200. Finally, the placement of EMG is more complex (electrodes need to be placed precisely with a target muscle). In contrast, IMU placement is simpler and involves placing the sensors on the finger phalanges. Our main goal behind this work is to use machine learning as a design material~\cite{ml_design_material} and enable designers to prototype custom microgestures without the need for having expertise in ML and programming. 

\subsection{How does this work differ from prior EMG toolkits in HCI?}
LibEMG~\cite{libEMG_ieee_access} and EMBody~\cite{embody_eics21} are EMG toolkits that enable novice users to develop EMG gesture recognition applications. Although an excellent starting point, EMBody’s~\cite{embody_eics21} contribution stands primarily in its open-access hardware design
and goal of enabling HCI researchers to explore EMG as an
input modality. EMBody’s~\cite{embody_eics21} limitation is that it was created with a predefined and rigid myoelectric control system. For example, by default, EMBody only supports a single classifier (SVM) and feature (root mean square). LibEMG~\cite{libEMG_ieee_access} aims to address these limitations of EMBody~\cite{embody_eics21} by providing users access to a richer set for feature selection and ML model parameters. The authors also tested this on multiple hardware platforms. However, these are limited to commercial EMG bands that do not offer the choice of optimizing the electrode placements or channels. 

In contrast with these exemplary prior works, we aim to understand how electrode selection schemes and classifiers can influence the accuracy of a given set of gestures. While the existing toolkits focus on easing the process of building ML models for a desired set of gestures and number of electrodes, our approach aims to identify a set of electrodes balancing sparsity and accuracy for a given set of gestures and desired set of electrodes.
% \subsection{IMU for gesture and activity detection}

%% file: design_goals.tex
\section{Design Goals}
Many toolkits and datasets exist for EMG gesture recognition across diverse disciplines, including biomedical engineering, prosthesis controls, HCI, and rehabilitation. Hence, the toolkits are often designed for specific tasks in a given discipline. For example, BioPatRec~\cite{ortiz2013biopatrec} is one of the very early toolkits designed in Matlab that enables researchers to interactively test and validate various ML features and algorithms suitable for prosthesis control. On the other hand, toolkits in HCI, such as LibEMG~\cite{libEMG_ieee_access} and EMBody~\cite{embody_eics21} focus on enabling novice users to quickly train an EMG recognition model with a custom gesture set defined by users. 

\subsection{Gesture Variability}
While existing EMG toolkits enable novice practitioners to create machine learning models for sEMG gesture recognition, they do not adequately support interaction designers in quickly prototyping and assessing the accuracy of specific gestures. To aid in gesture design prototyping, a primary goal of our design tool is to allow designers to choose from a set of gestures and to select and refine a subset of those gestures for effective recognition. \rev{This is a crucial requirement because, for a given set of $n$ gestures, there are $2^n -1$ subsets possible and being able to select and deselect gestures for given application scenarios rapidly helps in rapid design iterations. } The design tool should generate a sparse EMG layout based on user-defined electrode constraints for a given set of gestures.

\subsection{Supporting User-Defined Electrode Placement}
One of our primary design objectives is to support user-defined electrode placement. This feature enables researchers and practitioners to customize sEMG sensor layouts according to their specific experimental needs, anatomical constraints, or application requirements. Unlike rigid, predefined electrode configurations, our approach allows designers to specify their preferred electrode locations while ensuring that the resulting layout maintains both signal quality and spatial efficiency.

\subsection{Hardware Independence}
Another fundamental design objective which we consider is to ensure hardware independence, allowing solutions to scale seamlessly across various sEMG acquisition systems (e.g. EMG bands like Myo or traditional gel-electrode based sensing systems) available to researchers and practitioners. Given the diversity of EMG hardware, including differences in amplifier sensitivity, sampling rates, and electrode interfaces, the solutions generated should scale across multiple hardware configurations without fine-tuning hardware parameters or other technical modifications.

\subsection{Scalability Across Datasets}
EMG is one of the most active research areas explored by several research communities. Each paper has a new dataset with different gesture sets, hardware, model parameters, and participants. One of the design goals is to ensure that the tool can import a wide range of datasets, understand their semantics and gestures, and then allow users to explore all possible combinations with that dataset interactively. Additionally, the tool should also be able to allow for generating an optimal solution for a given set of parameters (e.g. gesture set, number of electrodes, specific muscles to choose from, ML parameters etc) that the user can pick from a chosen dataset. 

%% file: method_v3.tex
\section{Data-Driven Analysis for Identifying Sparse Configurations}

Electrode placement is crucial for robust gesture recognition, as different gestures activate distinct muscle groups \cite{7320073, hwang2014channel}. While uniform distribution around the forearm is common \cite{delta, grabmyo, putemg}, it often leads to suboptimal performance. Identifying the optimal configuration is challenging, especially with high-density arrays, due to combinatorial complexity $\binom{n}{k}$. To address this, we treat electrode selection as a feature subset selection problem, aiming to efficiently identify a minimal set of electrodes that maximizes accuracy while maintaining a sparse, practical layout.

\rev{\textbf{The goal of the data analysis is to identify the optimal combination of electrode selection schemes \rev{(ESS)} and classifiers \rev{(CL)} that balance accuracy, sparsity, and computational efficiency. This informs our design tool in generating sparse electrode layouts with high classification accuracy in real time.}}

\subsection{Method}
We systematically analyze electrode selection schemes and classifier combinations to identify the minimal subset of electrodes and evaluate their generalizability across six open-source datasets. Before diving into the analysis, we present the underlying methodology by introducing the electrode selection schemes, datasets, and the preprocessing pipeline.

\subsubsection{Experimental Datasets}
After conducting a literature survey based on three criteria: (1) forearm muscle coverage, (2) diversity of HCI-relevant gestures, and (3) open-access availability, we identified six relevant datasets, summarized in Table~\ref{tab:datasets}. For more details, readers can refer to the respective publications.

We begin our analysis with the CSL-HDEMG dataset~\cite{cslhdemg}, published at CHI 2015, for three reasons: (1) it includes 192 electrodes, offering extensive forearm coverage; (2) it aligns with HCI applications through single-handed gestures; and (3) it features 27 diverse hand gestures, grouped into tap, bend, and multi-finger sets.

The dataset was recorded from five users (five sessions/user). Each gesture was repeated 10 times per session, resulting in a total of
$27\ (\text{\# gestures}) \times 10\ (\text{\# trials}) \times 5\ (\text{\# users}) \times 5\ (\text{\# sessions}) = 6,750$ gesture trials.
Each gesture was performed over a 3-second interval, and the signals were recorded at a sampling rate of $2048$ Hz.

To empirically assess the generalizability of the electrode selection schemes, we also evaluate their performance across five additional datasets in Section~\ref{across_datasets}.

\subsubsection{Electrode Selection Schemes}

The feature subset selection problem aims to identify the most informative set of features, in this case, sEMG channels, that maximizes model accuracy. Broadly, subset selection methods fall into two categories: filter-based methods (model-independent, relying on data statistics) and wrapper-based methods (model-dependent, evaluating feature impact on model performance) \cite{kohavi1997wrappers}. In this paper, we evaluate four methods for selecting the most informative electrodes:

\begin{enumerate}[leftmargin=*]
    \item \textbf{Mutual Information (MI)}: Uses entropy-based calculations to measure the amount of information shared between features and target variables. By selecting electrodes based on their informational contribution, MI helps minimize redundant or irrelevant electrode placements. \cite{kraskov2004estimating,vergara2014review,electrode_configuration_ieee_tim_23}.
   
    \item \textbf{Permutation Importance (PI)}: It provides insight into the importance of each electrode by showing how much performance is affected when an electrode is removed.  This can help pinpoint the minimum number of electrodes necessary to maintain high accuracy while ensuring computational efficiency \cite{breiman2001random}.
   
    \item \textbf{Root Mean Square-based Importance (RMS-I)}: Quantifies muscular activity using RMS, which highly correlates with muscle force. \cite{de1997use}.
    
    \item \textbf{SHapley Additive exPlanations (SHAP)}: Uses game theory to assign importance scores based on feature contributions to predictions. It offers a robust explanation of feature importance and ensures that the most influential electrodes are selected, helping build a sparse yet highly accurate EMG layout  \cite{10.5555/3295222.3295230}.
   
\end{enumerate}

\subsubsection{Classifiers} \label{classifiers}
Rapid prototyping in HCI demands tools that balance speed, interpretability, and practical deployability. To systematically test our subset selection methods, we evaluate them in combination with six classifiers that are commonly used and well-established in EMG gesture recognition workflows: Support Vector Classifier (SVC), Logistic Regression (LR), K-Nearest Neighbors (KNN), Gaussian Naive Bayes (NB), XGBoost (XGB), and Random Forest (RF) with $max\_depth = 30$. All classifiers are used with their default hyperparameters. Additionally, we train a Multi-Layer Perceptron (MLP) pipeline as described in \cite{liu2022research} to explore the performance of neural networks in this context. We evaluated every electrode selection scheme and classifier combination, resulting in a total of 4 (\#ESS) $\times$ 7 (\#CL) = 28 combinations.

Each gesture trial is divided into three non-overlapping windows of equal length. For each window, we compute the Root Mean Square (RMS) value for each channel and concatenate the results, producing an $n \times 3$ feature vector per trial, where $n$ is the number of channels. This feature vector is used as input to each classifier.

\paragraph{\textbf{Other Reduction Strategies and Classification Techniques}}

SparseEMG prioritizes electrode reduction over general dimensionality reduction. Our primary objective is to identify and eliminate non-essential electrodes while preserving classification accuracy. Techniques like PCA and LDA are widely used for reducing data dimensions, but they don’t focus on electrode reduction, which is crucial for optimizing hardware efficiency. Similarly, methods like LASSO and autoencoders help with dimensionality reduction but don't directly address which electrodes can be discarded without sacrificing performance.

We included a simple neural network (MLP) in our evaluation to investigate the potential of deep learning. However, we deliberately excluded more complex architectures such as CNNs and LSTMs due to their high computational overhead, reliance on larger datasets, and reduced interpretability. SparseEMG emphasizes speed and interpretability, making simpler models like Random Forest and MLP more suitable for rapid prototyping and exploratory analysis.

While this paper focuses on applying existing machine learning techniques to electrode reduction, future work could explore the development of novel methods tailored to this challenge. Though beyond the scope of this study, these advancements represent an exciting direction for future research.

% \ad{see the alternative version above}
% \paragraph{\textbf{Why no Deep learning and Dimensionality Reduction?}}While dimensionality reduction techniques like PCA and LDA exist, they reduce data dimensions instead of directly focusing on electrode reduction, which is the primary goal. Similarly, methods like LASSO and autoencoders can reduce dimensions but do not explicitly target electrode reduction. For example, autoencoders compress data but do not directly eliminate less important electrodes. Although we did include a neural network(MLP), we excluded other deep learning networks due to their high computational complexity, need for larger datasets, and lack of interpretability. We chose the four electrode selection schemes for their balance of simplicity, interpretability, generalizability across datasets and gesture types, and computational feasibility. 

\subsection{Evaluating Selection Scheme and Classifier Combination} \label{analysis}
% \begin{tcolorbox}
% RQ\rev{1}: What is the minimum number of electrodes required for each ESS-CL combination to achieve maximum accuracy, and at what point does the accuracy plateau for each combination?
% \end{tcolorbox}

\vspace{4pt}
\noindent
\fbox{%
  \colorbox{gray!10}{%
    \parbox{\linewidth}{%
      \vspace{6pt}\hspace{6pt}%
      \parbox{\dimexpr\linewidth-12pt\relax}{%
        \textbf{RQ\rev{1}:} What is the minimum number of electrodes required for each ESS-CL combination to achieve maximum accuracy, and at what point does the accuracy plateau for each combination?
      }%
      \hspace{6pt}\vspace{6pt}
    }%
  }%
}
\vspace{4pt}

Previous works \cite{pelaez2022reducing, cslhdemg} have shown that not all electrodes contribute significantly to achieving maximum accuracy, and beyond a certain point, adding more electrodes results in minimal to no improvement. 
We evaluate each ESS-CL combination to determine the minimum number of electrodes required to reach its maximum possible accuracy, i.e., the point at which accuracy plateaus.

For this analysis, we included all the gestures from the dataset. Starting from the top two ranked electrodes, we progressively added more electrodes, up to a maximum of 100. This incremental strategy allowed us to determine the minimum number of electrodes required to achieve near-maximal classification performance. We included all five users from the dataset to evaluate generalization across different individuals. To account for variability across trials and sessions, we use a total of 20 trials across sessions per user, with 27 gestures, resulting in 540 trials per user. A 4-fold cross-validation is performed for each electrode selection scheme and classifier combination. During each cross-validation fold, the dataset is stratified to ensure a balanced distribution of gestures across the training and testing sets. In each fold, the test set contains 135 trials (25\%), with the remaining trials used for training. We repeat the entire cross-validation process across five randomly selected users from the dataset to evaluate the generalizability of our findings across different individuals.

\subsubsection{\textbf{Findings}}: The results in Table~\ref{tab:100_electrode_analysis} highlight the trade-off between the number of selected electrodes and the classification accuracy of different subset selection methods and classifiers. The PI + RF combination achieves the highest accuracy (94.04\%) with a relatively low electrode count (76), making it one of the most efficient combinations. Similarly, MI + RF (92.81\% accuracy, 78 electrodes) balances performance and efficiency, while MI + LR prioritizes minimal electrodes (74) at the cost of reduced accuracy (89.93\%). In contrast, methods like RMS-I and SHAP often demand significantly more electrodes for marginal gains. For example, SHAP + RF achieves only a 0.04\% accuracy increase over PI + RF (94.00\% vs. 94.04\%) but requires 94.40 electrodes, 18 more than PI + RF. Moreover, RMS-I + MLP achieves the highest overall accuracy (95.20\%) but at the expense of practicality, using 94.40 electrodes.
% \begin{tcolorbox}
% \textbf{Takeaway1}: PI + RF offers the best trade-off between accuracy and sparsity, achieving near-peak performance (94.04\%) with fewer electrodes(\#76). In contrast, methods like RMS-I and SHAP require substantially more electrodes for only marginal accuracy gains.
% \end{tcolorbox}
% \vspace{-4mm}
\begin{table}[!h]
  \centering
  \fontsize{6}{11}\selectfont
  \begin{tabular} {l | ll | ll | ll | ll}
    \toprule
    \textbf{Classifier} 
    & \multicolumn{2}{c|}{\textbf{MI}} 
    & \multicolumn{2}{c|}{\textbf{PI}} 
    & \multicolumn{2}{c|}{\textbf{RMS-I}} 
    & \multicolumn{2}{c}{\textbf{SHAP}} \\
    
   & \textbf{\# Elec.} & \textbf{Acc.} 
    & \textbf{\# Elec.} & \textbf{Acc.} 
    & \textbf{\# Elec.} & \textbf{Acc.} 
    & \textbf{\# Elec.} & \textbf{Acc.} \\
    \midrule

    \textbf{KNN} & 68 & 76.00 & 72 & 69.41 & 82 & 83.81 & 75 & 67.89 \\
    \textbf{LR}  & 74 & 89.93 & 91 & 89.33 & 82 & 90.78 & 92 & 89.07 \\
    \textbf{MLP} & 75 & 85.26 & 89 & 82.33 & 94 & 84.41 & 95 & 83.63 \\
    \textbf{NB}  & 85 & 79.78 & 93 & 81.33 & 71 & 84.52 & 88 & 81.93 \\
    \textbf{RF}  & 78 & 92.81 & 76 & 94.04 & 86 & 93.78 & 94 & 94.00 \\
    \textbf{SVC} & 48 & 49.19 & 90 & 30.41 & 85 & 83.74 & 96 & 32.04 \\
    \textbf{XGB} & 73 & 84.30 & 66 & 86.07 & 85 & 84.67 & 76 & 86.30 \\

    \bottomrule
  \end{tabular}
  \caption{Comparison of subset selection methods and classifiers for gesture recognition. For each method, the average number of selected electrodes (rounded off to the nearest integer) (\# Elec.) and best classification accuracy (Acc.) are reported.}
  \label{tab:100_electrode_analysis}
  \vspace{-6mm}
\end{table}

\subsection{Optimizing Electrode Count for Practical Usability}
% \begin{tcolorbox}
% RQ\rev{2}: How well do ESS-CL combinations perform when constrained to a practical electrode limit (e.g., 20 electrodes), and which combinations offer the best balance between accuracy and real-world hardware feasibility?
% \end{tcolorbox}

\vspace{4pt}
\noindent
\fbox{%
  \colorbox{gray!10}{%
    \parbox{\linewidth}{%
      \vspace{6pt}\hspace{6pt}%
      \parbox{\dimexpr\linewidth-12pt\relax}{%
        \textbf{RQ\rev{2}:} How well do ESS-CL combinations perform when constrained to a practical electrode limit (e.g., 20 electrodes), and which combinations offer the best balance between accuracy and real-world hardware feasibility?
      }%
      \hspace{6pt}\vspace{6pt}
    }%
  }%
}
\vspace{4pt}

Real-world wearable systems face tight hardware constraints, typically supporting no more than 16 electrodes\footnote{\url{https://tech.facebook.com/reality-labs/2021/3/inside-facebook-reality-labs-wrist-based-interaction-for-the-next-computing-platform/}}. \rev{Earlier, we found that PI and MI paired with RF and LR achieved high accuracy but required 75 to 80 electrodes, which is not practical for deployment}. We impose a constraint on the number of electrodes used during training to improve practicality. Anticipating future advancements, we set a slightly higher limit to accommodate hardware advancements, restricting our analysis to top 20 electrodes.

\subsubsection{\textbf{Findings}}
\begin{figure*}
    \centering
    \includegraphics[width=1\linewidth]{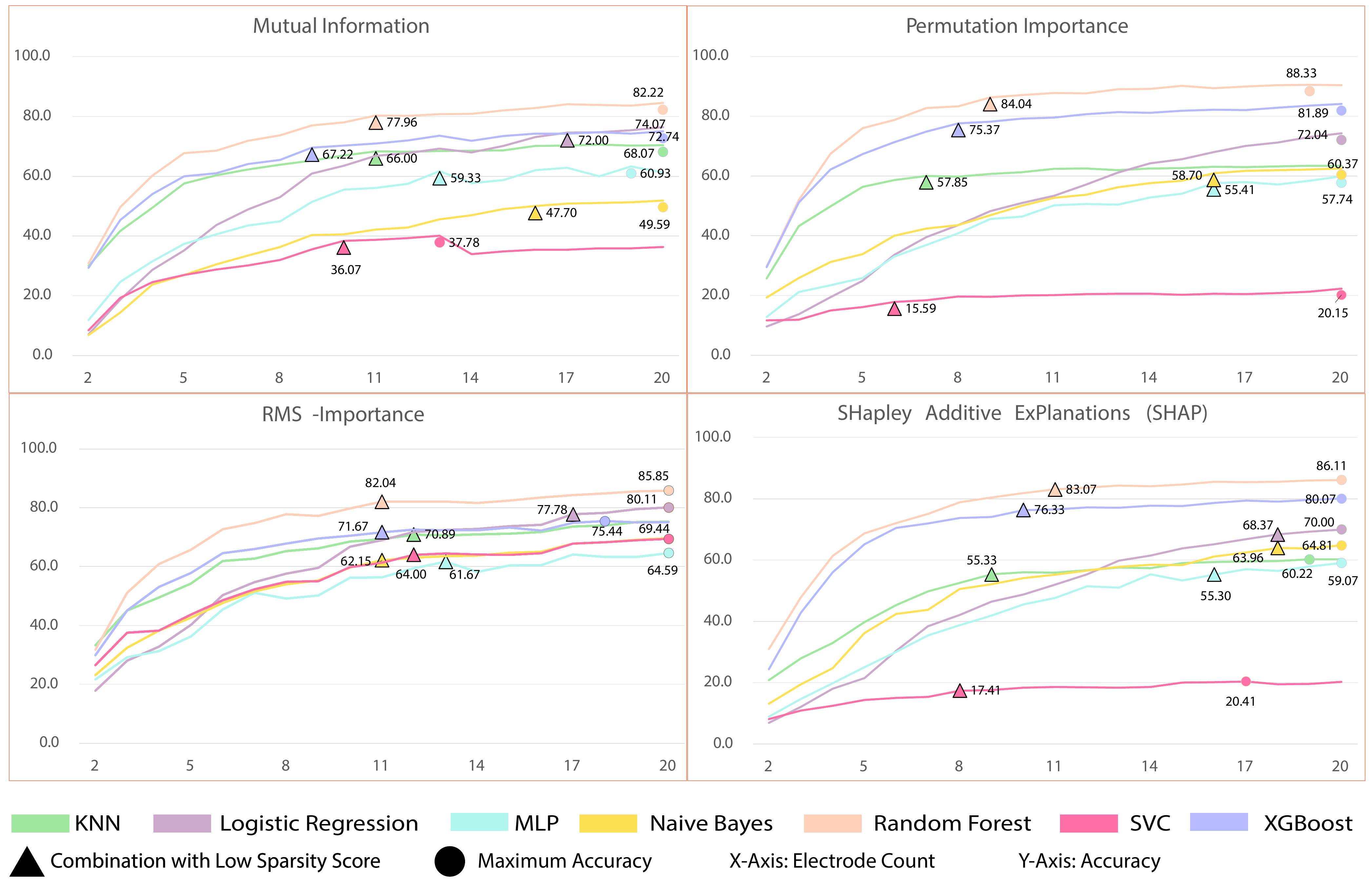}
    \caption{\rev{Comparison of Electrode selection and classifier combinations for a constraint of 20 electrodes across five users and 29 gestures. For each of the combinations, the maximum accuracy and the optimal configuration derived from sparsity scores are highlighted respectively.}}
    \label{fig:20_electrode_combination}
\end{figure*}

: Figure \ref{fig:20_electrode_combination} shows the accuracy trends across five users, averaged for each electrode count, and highlights the plateau behaviour with a limited number of electrodes. In particular, RF performs well across all selection methods, achieving 82\%–88\% accuracy within 20 electrodes. Among these, the PI + RF combination achieves the highest accuracy, reaching 88.33\% with 19 electrodes. In contrast, SVC struggles to achieve high accuracy, even when using all 20 electrodes across various selection methods.

% \begin{tcolorbox}
% \textbf{Takeaway2}: PI + RF delivers the highest accuracy (88.33\%) even with a constraint of 20 electrodes, demonstrating strong performance under hardware-constrained conditions, while SVC consistently underperforms even with the full set of 20 electrodes.
% \end{tcolorbox}

\subsection{Balancing Sparsity and Accuracy}\label{sec:sparsity_accuracy}

% \begin{tcolorbox}
% \end{tcolorbox}

\vspace{4pt}
\noindent
\fbox{%
  \colorbox{gray!10}{%
    \parbox{\linewidth}{%
      \vspace{6pt}\hspace{6pt}%
      \parbox{\dimexpr\linewidth-12pt\relax}{%
        \textbf{RQ\rev{3}:} Which combinations offer the best balance between classification accuracy and sparsity?
      }%
      \hspace{6pt}\vspace{6pt}
    }%
  }%
}
\vspace{4pt}

Our previous analysis (Figure~\ref{fig:20_electrode_combination}) showed how each combination performed across users, gestures and sessions with a constrained set of 20 electrodes. Figure~\ref{fig:20_electrode_raw_data} shows the raw data of the accuracies for each of the combinations across each electrode count. Based on this data, selecting the best combination solely based on the lowest electrode count and highest accuracy may not be optimal, as some combinations exhibit early plateaus. For instance, SHAP and RMS-I show diminishing returns, where adding electrodes beyond a certain threshold provides negligible improvements. In these cases, the trade-off between additional electrodes and minimal accuracy gains becomes unfavorable. A notable example is MI + SVC, which achieves its peak accuracy using only 13 electrodes. However, the accuracy remains low at 37.78\%. This illustrates the risks of prioritizing electrode reduction without considering classifier compatibility and emphasizes the importance of analyzing the trade-off curve rather than focusing solely on extreme values.

To address this, we develop a selection metric called ``\textbf{Sparsity Score}', a metric designed to evaluate and balance the trade-off between the number of electrodes used and the classification accuracy achieved by a specific ESS-CL combination. It quantifies the point at which adding more electrodes provides diminishing returns in accuracy, penalizing combinations that require significantly more electrodes to achieve only marginal improvements in performance. The ``\textbf{Sparsity Score}' is calculated as:

$Sparsity~Score(E) = (w_1 * (100 - Accuracy(E)) + (w_2 * E) $, where, $(w_1 + w_2 = 1)$

We used a linear weighted scheme to balance the accuracy drop and the number of electrodes; however, an exponential scheme could also be used to assign steeper penalties for accuracy drop or electrode count. A lower sparsity score indicates a better balance between sparsity and accuracy. For this specific analysis, we set the weights equally (i.e $w_1 = w_2 = 0.5$).
We determine the plateau point for each combination by identifying the electrode count $E$ that minimizes this score. The scores are then averaged across the five users, and the combination with the lowest overall score is selected.

\subsubsection{\textbf{Findings}}: The annotated triangles in Figure~\ref{fig:20_electrode_combination} demonstrate the sparsity scores for a given electrode count and accuracy. For e.g. for PI + RF, the best accuracy is 88.33\%, but the best sparsity score is achieved at 9 electrodes, achieving an accuracy of 84.04\%. Similarly, for PI + SVC, the sparsity scores are always higher because the accuracies are low. Despite this, the Sparsity Score enables us to identify a solution with a 70\% drop in electrode count while having only a $\sim$4.5\% drop in accuracy. This approach ensures a more practical and efficient electrode selection strategy. A key finding from this analysis is that PI + RF is the best combination for this dataset. However, it is essential to note that this conclusion is based on results from a single dataset and cannot be generalized across multiple datasets.

\subsection{Generalizability of the Selected Method Across Datasets} \label{across_datasets}

\vspace{4pt}
\noindent
\fbox{%
  \colorbox{gray!10}{%
    \parbox{\linewidth}{%
      \vspace{6pt}\hspace{6pt}%
      \parbox{\dimexpr\linewidth-12pt\relax}{%
        \textbf{RQ\rev{4}:} SHAP + RF and PI + RF offer the best balance between classification accuracy and sparsity. But how well do they generalize across datasets?
      }%
      \hspace{6pt}\vspace{6pt}
    }%
  }%
}
\vspace{4pt}

To ensure the robustness of the proposed method, the selected combination must generalize well across diverse datasets, including potential custom datasets. To evaluate this generalizability, we extend the analysis from Section \ref{sec:sparsity_accuracy}, where PI + RF combination demonstrated the best balance between sparsity and accuracy, to the additional datasets listed in Table \ref{tab:datasets}.

\begin{table}[!h]
  \centering
  \fontsize{7}{11}\selectfont
  \begin{tabular}{ l c c l }
    % p{0.2\linewidth} 
    % p{0.14\linewidth} 
    % >{\centering}p{0.14\linewidth} 
    % >{\centering}p{0.14\linewidth}
    % p{0.22\linewidth}}
    
    % \toprule
%     & & \multicolumn{2}{c}{\small{\textbf{Test Conditions}}} \\
%     \cmidrule(r){3-4}
    {\small\textbf{Dataset}}
    % & {\small\textbf{Setup}}
    & {\small\textbf{Channels}}
    & {\small\textbf{Gestures}}
    & {\small\textbf{Gesture Types}}
     \\

    \toprule   
      CSL-HDEMG~\cite{cslhdemg} & 192 & 29 & Single-finger, Multi-finger \\ 

      Hyser~\cite{hyser} & 256 & 34 & Single-finger, Multi-finger, Wrist \\ 

      DELTA~\cite{delta} & 64 & 6 & Multi-finger, Wrist \\ 
      Nizamis et al.~\cite{nizamis2020characterization} & 64 & 7 & Single-finger, Wrist \\

      GrabMyo~\cite{grabmyo} & 28 & 17 & Single-finger, Multi-finger, Wrist \\ 

      putEMG~\cite{putemg} & 24 & 8 & Multi-finger, Wrist \\ 
    \bottomrule
  \end{tabular}
 \caption{ Datasets used for identifying the most optimal combination of subset selection method and classifier}~\label{tab:datasets}
 \vspace{-2em}
\end{table}

Each dataset varies regarding the number of users, gestures, electrode hardware, sessions, and placement strategies. To ensure consistency across datasets, we took the following steps: (1) we selected the number of users that was the least common denominator across all datasets, and (2) we chose 20 trials per dataset, spread across sessions, to account for session variability.

We trained all combinations of subset selection methods and classifiers using data from five randomly selected users per dataset. Each combination was evaluated using 4-fold cross-validation. For each fold, the data was split in a stratified manner into training (75\%) and testing (25\%) sets, with the following splits: Hyser~\cite{hyser} (306 train, 102 test), DELTA~\cite{hyser} (90 train, 30 test), Nizamis et al.~\cite{nizamis2020characterization} (105 train, 35 test), GrabMyo~\cite{grabmyo} (255 train, 85 test), and putEMG~\cite{putemg}
(105 train, 35 test).

For each user in a dataset, the Sparsity Score is computed for all combinations, and the average score across five users is then calculated. These scores are further averaged across datasets to determine the final score for each combination.

\subsubsection{\textbf{Findings}}
\begin{figure}[!h]
    \centering

    \includegraphics[width=\linewidth]{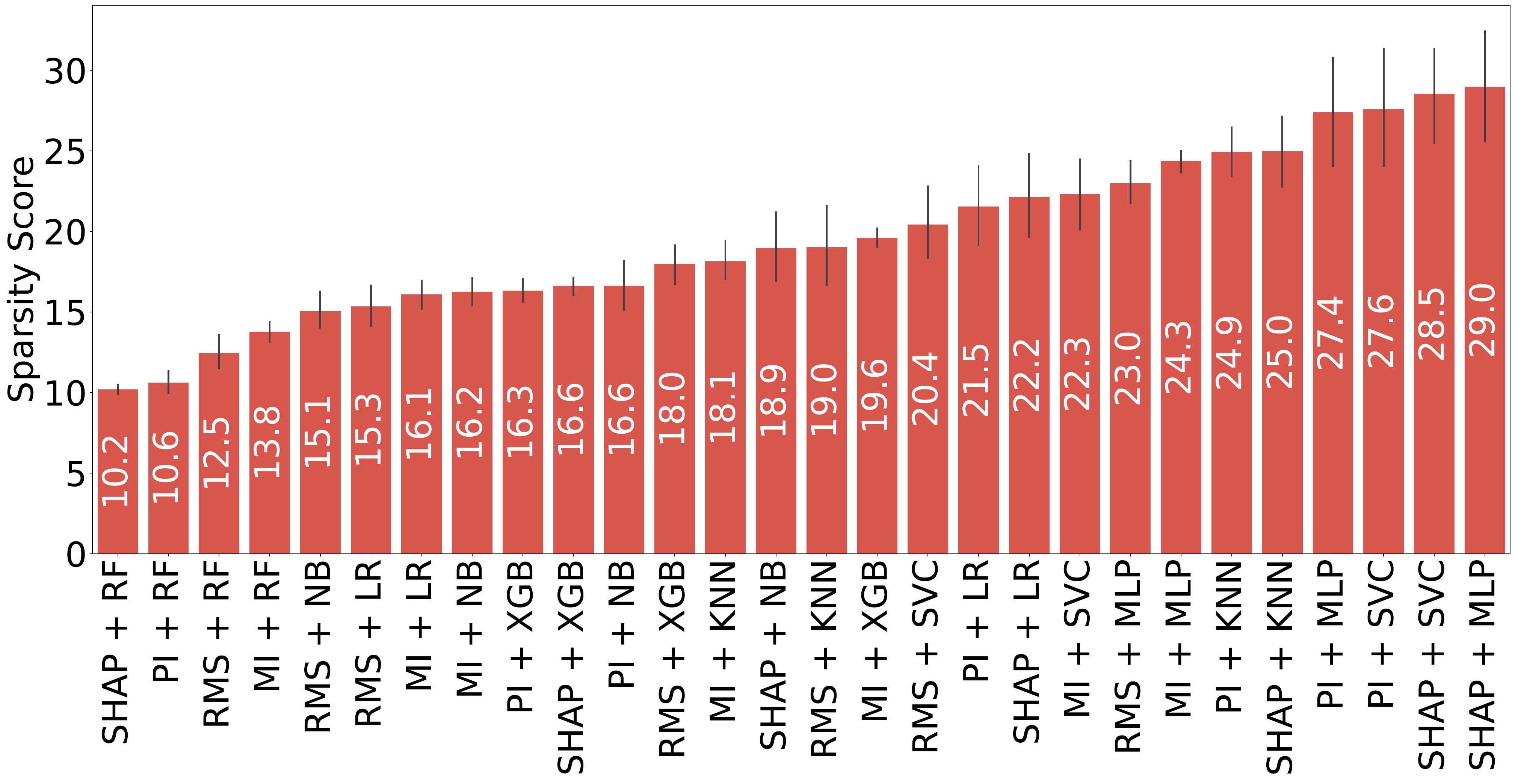}
    
    \caption{Average combination scores of subset selection methods with classifiers across all six datasets listed in Table \ref{tab:datasets} and five random users. Lower scores indicate better performance, with Permutation Importance (PI) + Random Forest (RF) achieving the lowest scores}
    \label{fig:averaged_combination_scores}

\end{figure}

Figure~\ref{fig:averaged_combination_scores} presents the averaged \textsc{Sparsity Score} across all datasets. The results show that the SHAP + RF combination performs the best, with the lowest score of 10.2. Interestingly, the PI + RF combination also achieves a similar score of 10.6. These are followed by RMS + RF (12.5), MI + RF (13.8), RMS + NB (15.1), and RMS + LR (15.2), all demonstrating relatively strong performance.

In contrast, SVC and MLP consistently emerge as the weakest classifiers with the added constraint on the number of electrodes, leading to higher sparsity scores across their respective combinations.

\subsection{Computation Time}

\vspace{4pt}
\noindent
\fbox{%
  \colorbox{gray!10}{%
    \parbox{\linewidth}{%
      \vspace{6pt}\hspace{6pt}%
      \parbox{\dimexpr\linewidth-12pt\relax}{%
        \textbf{RQ\rev{5}:} What is the variation in computation time across different electrode selection methods and classifiers?
      }%
      \hspace{6pt}\vspace{6pt}
    }%
  }%
}
\vspace{4pt}

One key criterion for the SparseEMG design tool is its ability to generate sparse layouts in near real-time, enabling rapid prototyping and iteration for gesture design. Since SHAP and Permutation Importance (PI) achieved the best Sparsity Scores across all datasets, it is essential to evaluate their suitability for real-time computation. For this, we analyzed the computation times for each electrode selection schemes (PI, MI, RMS-I, and SHAP) on the first dataset (CSL-HDEMG~\cite{cslhdemg}), across users and for a randomly chosen session with 27 gestures. MI and RMS-I are filter-based, model-independent methods, so their computation times were consistent across users. Computation times were recorded on a desktop machine (Intel Core i7 [13th Generation], 32GB RAM, and NVIDIA GTX-4060 GPU).

\subsubsection{\textbf{Findings}}
\begin{figure}
    \centering

    \includegraphics[width=1\linewidth]{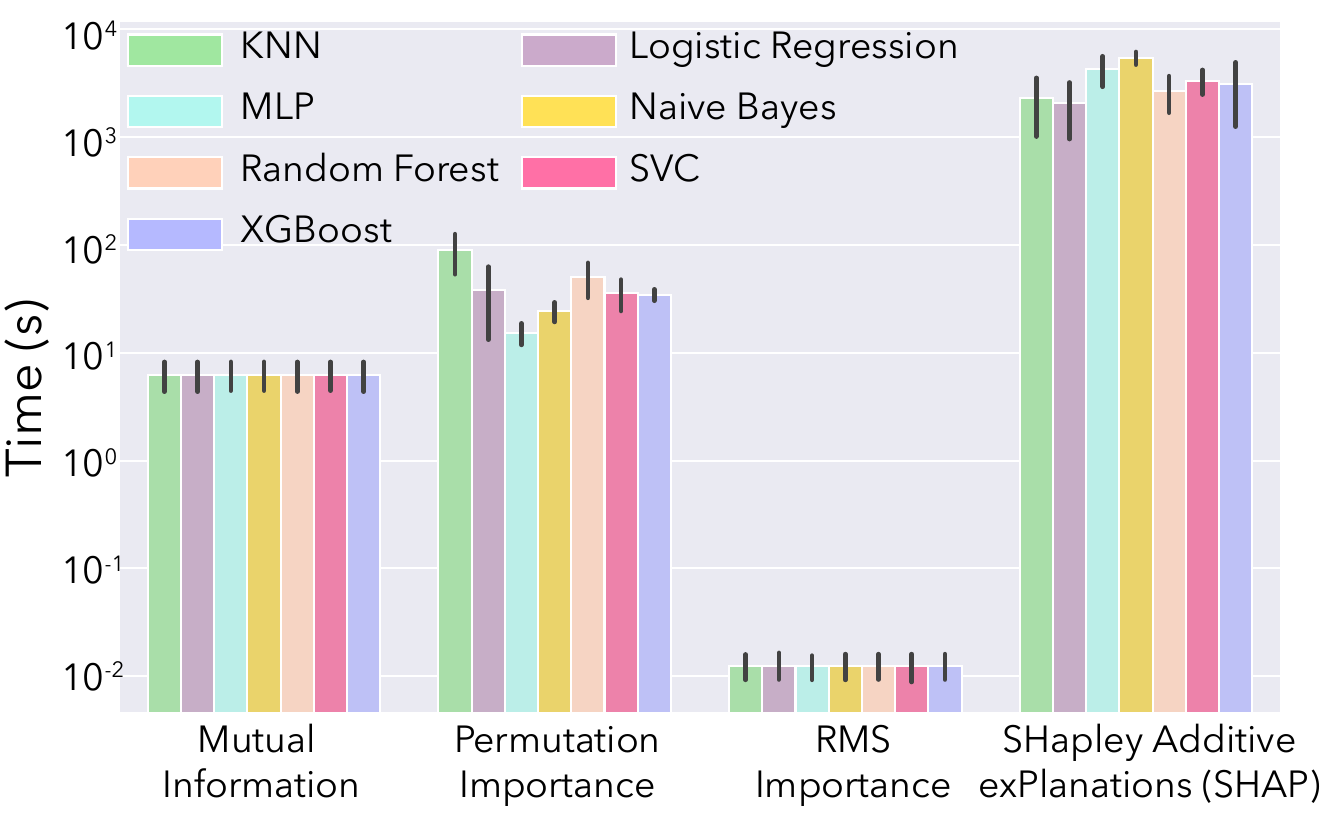}
    \caption{\rev{Computation times for the four electrode selection schemes.}}
    \label{fig:computation_times}
    
\end{figure}

Figure~\ref{fig:computation_times} illustrates the computation times for each selection method across all classifiers and users. SHAP took the longest time, averaging 3322.48 $\pm$ 1097.4 seconds, while RMS-Importance was the fastest, taking only 0.012 $\pm$ 0.003 seconds. Given the significant computational cost of SHAP, we excluded it from the design tool and selected the next best combination, Permutation Importance + Random Forest, as the primary electrode selection-classifier combination.

%% file: design_tool.tex
\subsection{Summary of Findings}
The key takeaways from the above in-depth analyses are:

\begin{enumerate}
    \item PI + RF offers the best trade-off between accuracy and sparsity, achieving near-peak performance (94.04\%) with fewer electrodes (\#76). In contrast, methods like RMS-I and SHAP require substantially more electrodes for only marginal accuracy gains.

    \item  PI + RF delivers the highest accuracy (88.33\%) even with a constraint of 20 electrodes, demonstrating strong performance under hardware-constrained conditions, while SVC consistently underperforms even with the full set of 20 electrodes.

    \item SHAP + RF and PI + RF consistently perform well in providing an optimal balance between sparsity and accuracy.

    \item SHAP's high computation time led to choosing PI + RF. 
\end{enumerate}

\section{Computational Design Tool for Rapid Selection Of Custom Sparse Layouts } \label{sec:tool}

\begin{figure*} [!h]
    \centering
    \includegraphics[width=1\linewidth]{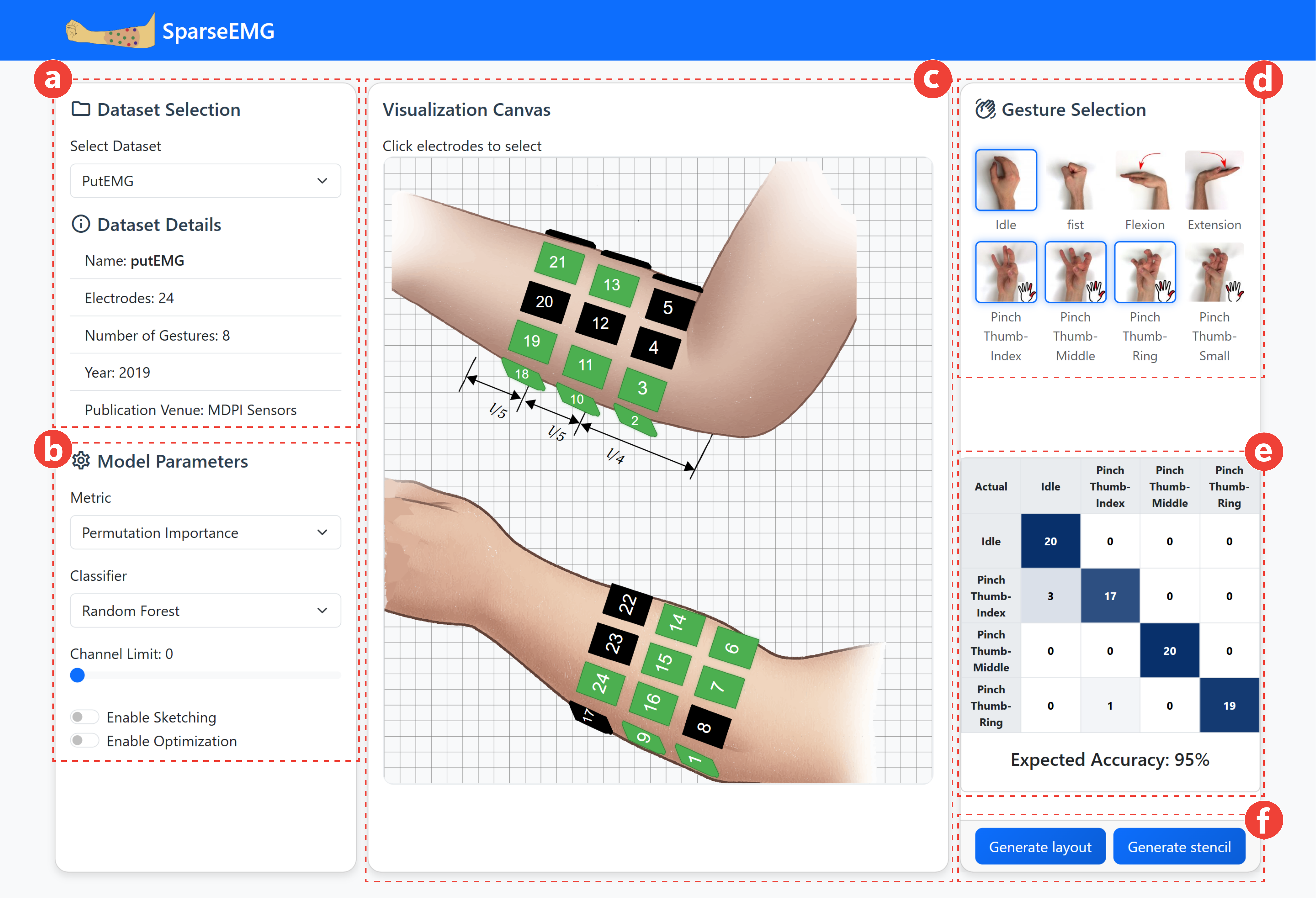}
    \caption{Screenshot of the SparseEMG Computational Design Tool}
    \label{fig:tool_screenshot}
\end{figure*}

Based on our analysis, we contribute a computational design tool. It assists designers in the following tasks.

\begin{itemize}[leftmargin=*]
    \item \textit{Finding a sparse EMG layout that balances high gesture recognition accuracy and sparsity:} Using the designer’s specifications, the tool selects optimal designs in near real time and indicates the expected recognition accuracy. This also allows the designer to quickly understand how well a desired set of gestures can be recognized with a specified set of electrodes. The design tool assists designers in locating the electrode locations and 

    \item \textit{Exploring location alternatives:} Considerations of ergonomic wearability or aspects inherent to certain application cases may restrict the space where sEMG electrodes can be deployed on the user’s forearm. For instance, an EMG band (e.g. MYO) with a pre-defined set of channels can only be placed on the upper region of the forearm. Similarly, an application case involving forearm actions may benefit from electrodes placed on the middle or upper region of the forearm rather than close to the fingertips. The tool allows the designer to select what locations can be augmented with sEMG electrodes and to explore alternatives quickly (Figure~\ref{fig:tool_screenshot}(c)). 

    \item \textit{Fine-tuning the Gesture Set for a given Electrode Configuration}: One key functionality of the design tool is to provide a visual representation that depicts the performance of the individual gestures (Figure~\ref{fig:tool_screenshot}(e)). This enables the designer to quickly inspect which gestures perform well and which do not and choose the most compatible gestures with high recognition accuracy. 

    \item \textit{Assisting in Electrode Placement:} One of the key challenges in EMG gesture recognition is the placement of electrodes. Replicating a given electrode placement is hard and can lead to significant performance issues~\cite{nittala2021computational,embody_eics21}. To assist the designers in replicating electrode placement for a given dataset, our design tool generates a placement stencil based on the details provided in the dataset.

\end{itemize}

\subsection{Walkthrough}
A screenshot of the design tool is shown in Figure~\ref{fig:tool_screenshot}. The designer first selects a dataset from a drop-down list. Once selected, the dataset details are loaded, including the gestures in the dataset, illustrations showing electrode placement. Next, she chooses the set of gestures that shall be recognized and selects the electrodes. The electrodes can be chosen by either directly selecting the electrodes or by sketching a region on the canvas. Then, the designer can place additional constraints for electrode placement. For instance, the designer selects the desired number of electrodes using the slider in the 'Model Parameters' panel.
Additionally, the user can choose the electrode selection mechanism and the desired classifier. The parameters are sent to the backend with the click of a button. To visually present the recognition accuracy of chosen gestures, the tool displays a confusion matrix and the location of the individual electrodes on the canvas. If the designer is unsatisfied with the Tool’s recommendation, she can quickly explore options iteratively. For instance, she may fine-tune the gestures or explore alternative locations for placing electrodes. \rev{We provide the source code to use our design tool \footnote{https://github.com/difflab-ucalgary/SparseEMG}}.

\subsection{Implementation}
 Notably, our tool differs from a conventional lookup table, which would require more than 4.2 billion entries for just one dataset (24 electrodes and 8 gestures) to cover the various combinations of electrodes and subsets of gestures. The front end of the design tool is implemented in  HTML/JavaScript and uses the Snap.svg library.
 
 \subsubsection{Dataset Scalability} To ensure easy and seamless integration of the datasets, we meticulously curated JSON files to add details of the dataset, such as the number of gestures, illustrations of gestures, electrode placement and other metadata (e.g. authors, electrode dimensions, spacing, etc.). For each dataset, we recreated the placements based on the instructions and descriptions from the publications. To allow the selection of electrodes within the tool, we created SVG files of the electrode placement and labeled them with their IDs. Once the user selects an electrode, we retrieve its ID through the Snap.svg library and update the global list, which consists of data about all the electrodes, their locations, and their selection status. We employ a similar approach for gesture selection as well.

\subsubsection{Solution Generation}

The backend of the tool is built using Python and the FastAPI framework. It receives information such as the dataset name, selected channels, maximum number of channels, gestures to recognize, subset selection method, and classifier type from the frontend via a WebSocket API call. The backend processes this information and begins training models using the top 2 to a maximum of 20 electrodes (or using the maximum number of channels, or only the selected channels). It performs 4-fold cross-validation for each electrode count to find the optimal sparse electrode configuration. After identifying the sparse configuration, the backend calculates the expected accuracy and confusion matrix for the selected gestures and sends the model file, accuracy, and confusion matrix back to the frontend.
 
 Finally, the front-end receives this information, highlights the sparse electrode configuration, plots the confusion matrix, and displays the predicted accuracy.

\subsection{Stencil Generation}

The tool also offers the flexibility to generate a stencil based on the user's arm measurements (e.g., the arm's circumference at various levels as described in the original paper of the selected dataset). The placement stencil is created based on the electrode size and spacing dimensions reported in the chosen dataset.  This stencil can be passed to any cutting machine. Once cut, the electrodes can be easily positioned on the user's arm according to the suggested sparse configuration.

\begin{figure} [!h]
    \centering

    \includegraphics[width=1\linewidth]{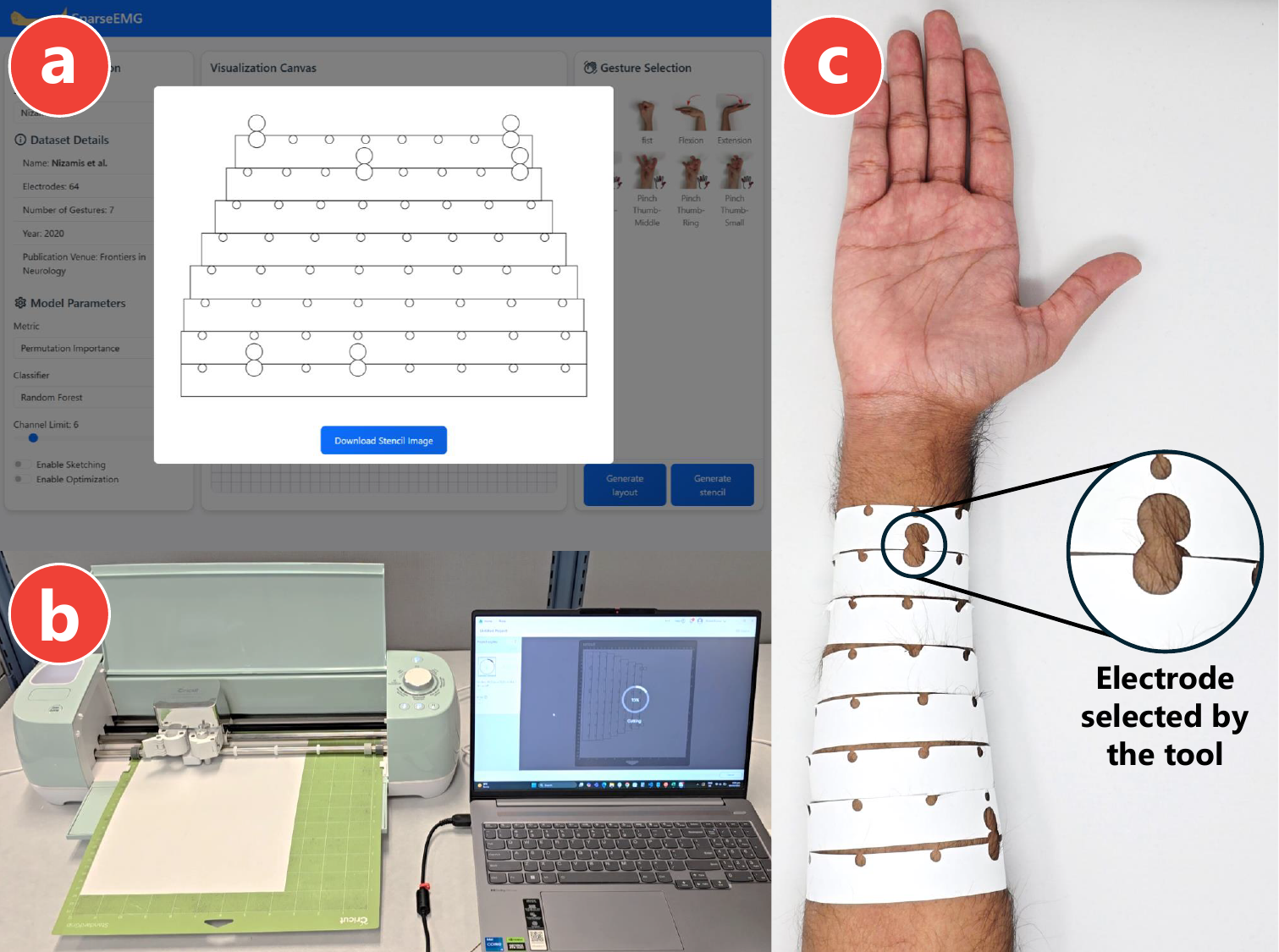}
    \caption{Applying electrode layout stencils generated by SparseEMG: (a) User interface for downloading the suggested sparse layout; (b) cutting the stencil using a paper cutter; (c) applying the stencil to the user’s forearm for consistent and repeatable electrode placement.}
    \label{fig:stencil_generation}
    
\end{figure}

%% file: applications.tex
\section{Does the Placement Suggested by S\MakeLowercase{parse}EMG Scale Across Users?}

\begin{figure*}[!ht]
    \centering
    \includegraphics[width=1\textwidth]{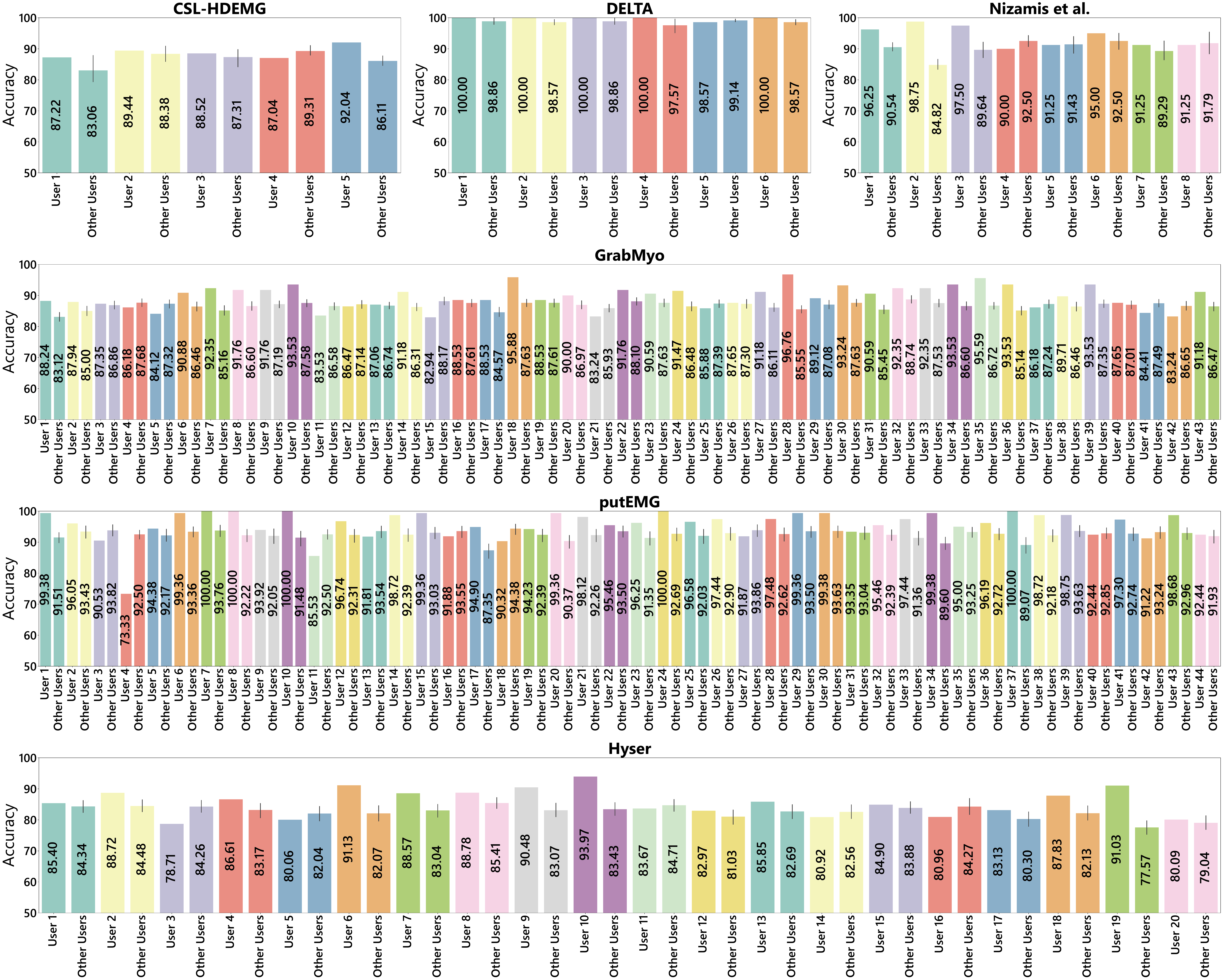}
    \caption{Cross-user validation accuracy for six datasets. Each sub-plot shows the accuracy for individual users using their own electrode layout (left bar) and the averaged accuracy when applying the user's electrode layout to other users (right bar)}
    \label{fig:combined_accuracy_plots}
\end{figure*}

One of the main goals of the tool is to reduce the number of electrodes without needing to individually optimize placement for each new user. For practical deployment, it is essential that the electrode configurations identified by the tool generalize well across different users. If a sparse configuration derived from one user can be applied to others with minimal or no loss in performance, it indicates that the layouts generated by the tool are scalable across unseen users. 

% \begin{tcolorbox}
%     RQ: Does the placement suggested by the tool scale across users in the same dataset, i.e. how does a placement layout for one user (suggested by PI+RF) perform on other users?
% \end{tcolorbox}

\subsection{Method}
We evaluated the tool using cross-user validation within each dataset. For each dataset, we selected one user to determine the optimal sparse electrode configuration based on their data. We then applied this configuration to the remaining users in the dataset and measured classification accuracy. This process was repeated for all the users, ensuring each served as the source-user once.

We applied this protocol across all six datasets to assess consistency and evaluate how well the tool's suggested placements generalized to other users. This validation helped determine if the tool identifies electrode placements that are robust and transferable across datasets.

% To evaluate this, we performed cross-user validation within each dataset. Specifically, for a given dataset, we selected one user and used the tool to determine the optimal sparse electrode configuration based solely on that user's data. This configuration was then applied to four other users from the same dataset, and classification accuracy was measured. We repeated this process for each of the five users, ensuring that each served as the source-user once.

% The same evaluation protocol was applied across all six datasets to ensure consistency and to assess how well the tool’s suggested placements scaled to other users. This cross-user validation allowed us to determine whether the tool identifies electrode placements that are robust and transferable within datasets.

\subsection{Results}
Figure~\ref{fig:combined_accuracy_plots} displays the results of the cross-user validation across all six datasets. Each plot shows a bar representing the accuracy for each user using their own electrode placement, alongside a bar showing the averaged accuracy when this user's layout is applied to others.

The tool-generated layouts demonstrated strong generalizability. The key takeaway is not the absolute accuracy but the consistency of accuracy when a layout from one user is applied to others. Ideally, if accuracy remains stable across users, it suggests that the tool generates configurations that generalize well. Datasets like CSL-HDEMG, DELTA, and Hyser exhibit minimal accuracy variation (2–6\%), while putEMG shows slightly higher variability, with performance differences reaching 6–8\% depending on the source layout.

These findings confirm the tool’s ability to identify robust, transferable electrode layouts that scale without requiring per-user calibration. However, since this analysis used users within the same dataset with consistent conditions, we conduct an application-driven analysis in section \ref{sec:application_driven_validation} to test the effectiveness of the tool's layouts across different users and hardware setups.

\rev{

\begin{figure*}[!ht]
\centering
\includegraphics[width=1\textwidth]{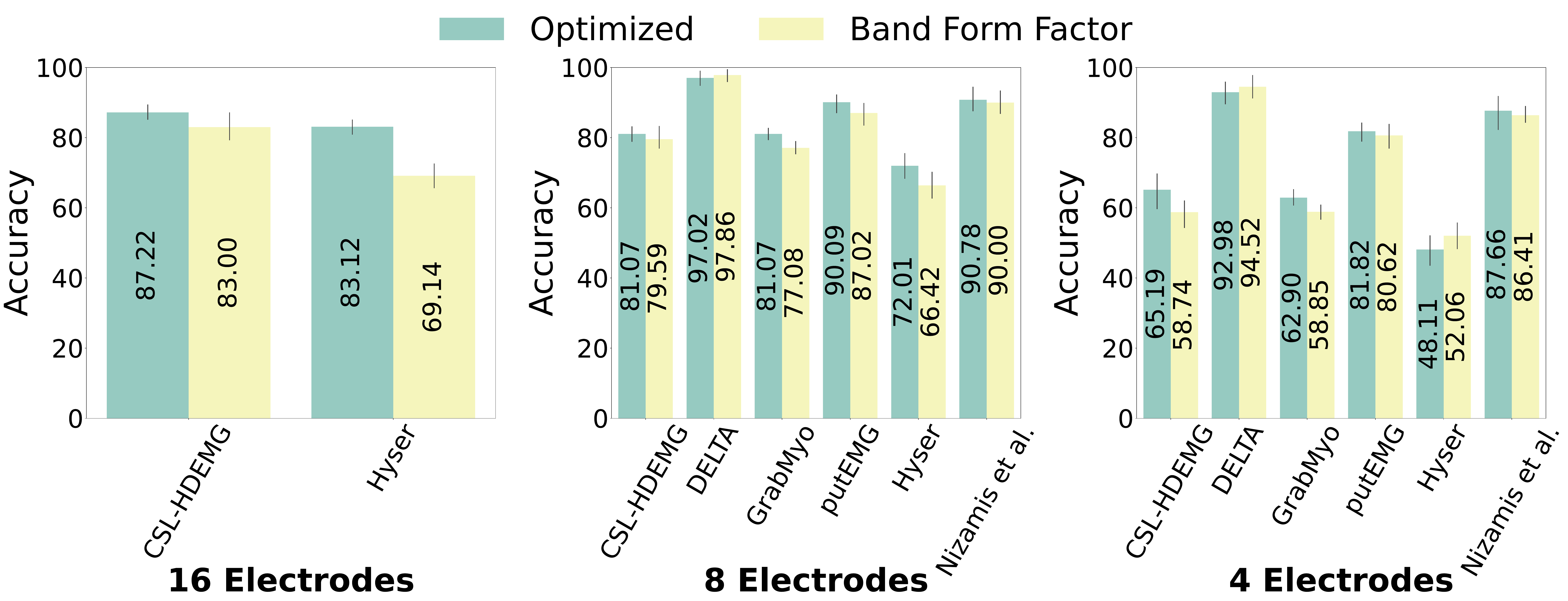}
\caption{Comparison of classification accuracy using SparseEMG-selected electrodes (left bars) versus standard band form-factor layouts (right bars) across three electrode configurations (16, 8, and 4 electrodes) and six datasets. For each configuration, the number of electrodes in the band layout was matched to that of SparseEMG for fair comparison. The 16-electrode configuration was evaluated only on the CSL-HDEMG \cite{cslhdemg} and Hyser \cite{hyser} datasets, as these were the only datasets with sufficient electrodes to support a circular band layout of that size.}
\label{fig:standard_vs_sparse}
\end{figure*}
}
% As shown in Figure ~\ref{fig:standard_vs_sparse}, the SparseEMG-selected electrode sets consistently outperformed the standard band form-factor layout across all datasets. The performance gain was particularly prominent in datasets with high electrode density, such as CSL-HDEMG \cite{cslhdemg} and Hyser \cite{hyser}, where the optimal electrode locations varied significantly from uniform spacing.

% SparseEMG improved the classification accuracy by 2–14\% compared to the band layout. This confirms that intelligent electrode selection can lead to more efficient and accurate gesture recognition without increasing hardware cost or electrode count.

% These results highlight the advantage of adaptive, data-driven electrode placement strategies over traditional uniform band-based designs, especially in applications where performance is critical and the number of electrodes must be minimized.

% \\

% ALTERNATIVE -- NEEDS INPUT

\rev{
\section{S\MakeLowercase{parse}EMG vs. Band Form-Factor}

Commercial EMG systems often employ fixed, evenly-spaced band configurations for convenience. In contrast, SparseEMG identifies performance-optimal sparse electrode subsets without geometric constraints. This section compares the classification accuracy of SparseEMG-generated layouts against typical fixed-band layouts, using identical electrode counts to assess the performance benefits of data-driven placement for interactive systems.

\subsection{Method}

We evaluated three standard band form-factor layouts (16, 8, and 4 electrodes), mimicking commercial systems by selecting equally spaced electrodes. These were directly compared against SparseEMG-derived layouts with the same electrode counts. The 16-electrode band was applied only to CSL-HDEMG \cite{cslhdemg} and Hyser \cite{hyser} datasets due to their suitable circular electrode arrangements. To generate the band-form factor designs, we used our computational tool (Section \ref{sec:tool}) to manually select the electrodes that form a band around the upper forearm. For the SpaseEMG layouts, we used our design tool. We set the channel count to 16, 8 or 4 appropriately to generate the optimized electrode placement.  Both the configurations (Band form-factor and SparseEMG layouts) used the same classifier (Random Forest, default parameters) and train-test split (75-25\%).

\subsection{Results}

As shown in Figure~\ref{fig:standard_vs_sparse}, SparseEMG consistently outperforms fixed band layouts across all datasets and electrode counts. On average, SparseEMG improves classification accuracy by $9$ percentage points at 16 electrodes, $3$ points at 8 electrodes, and $1$ point at 4 electrodes. These gains are particularly pronounced in low-electrode settings, where the band layout often falls below 60\% accuracy, while SparseEMG maintains significantly higher performance. For instance, on the CSL-HDEMG \cite{cslhdemg} dataset, accuracy with four band-placed electrodes drops to 58.74\%, whereas SparseEMG retains 65.19\%, yielding a 7-point improvement. This trend holds across nearly all datasets, highlighting SparseEMG's robustness under sensor constraints and its potential to substantially enhance performance in wearable EMG systems where electrode count is limited. }

\section{Application-Driven Validation} \label{sec:application_driven_validation}
A thorough usability evaluation of our computational design tool is challenging because of the multi-dimensional space involving electrode count, placement, datasets, gestures, and ML parameters resulting in combinatorial complexity. As recommended by Ledo et al., we adopt application-driven validation~\cite{toolkit_evaluation_led_chi18} to evaluate the practical usefulness of our tool. This approach has also been commonly employed in prior gesture \rev{work}.
% \begin{tcolorbox}
%     RQ: Does the placement suggested by the tool scale to real-world applications using completely different hardware setups, i.e. how does a placement layout for one user (suggested by PI+RF) perform on other users?
% \end{tcolorbox}

Our approach draws inspiration from prior work that contributed computational design tools~\cite{sparseIMU_tochi23, wrlkit_uist23}. We collected three datasets based on different application scenarios with varying hardware configurations and participants to assess the tool's practical usefulness and generalizability to real-world applications. This section compares the accuracy achieved using the channels specified in the layouts suggested by the tool for the original dataset, as well as when these layouts are applied to participant recordings.

\subsection{Acquisition Setup} We chose widely available recording hardware to make the setup more practical. Since Arduino-based hardware is commonly used in EMG gesture recognition, we selected the Olimex EMG Shield for Arduino-based boards to collect data for two datasets. This hardware allows stacking multiple shields and supports data collection from up to six channels with a single Arduino. We used six channels to record the data. The sampling rate is 256 Hz, the maximum supported by the board.

\begin{figure} [!h]
    \centering

    \includegraphics[width=1\linewidth]{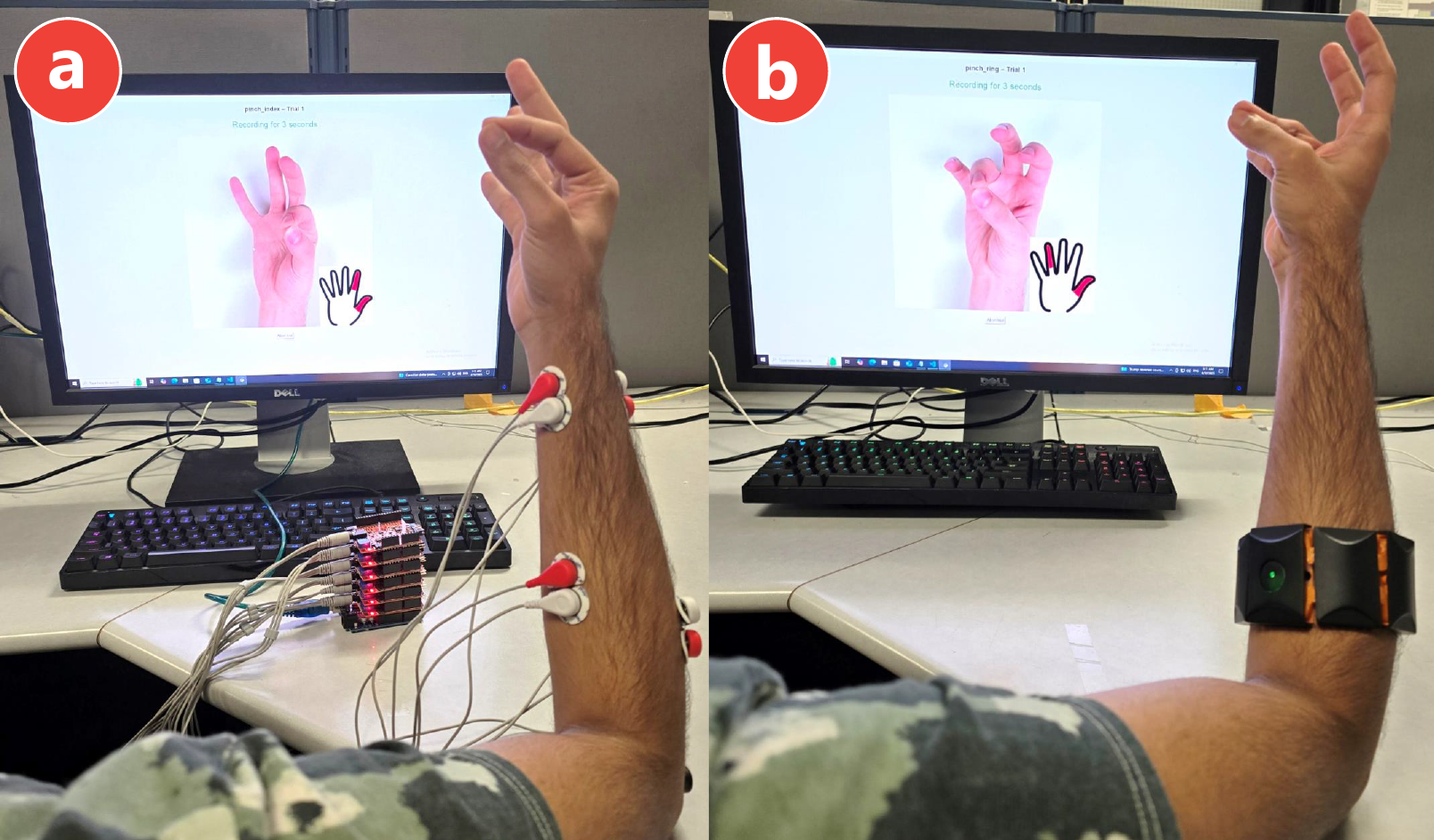}
    \caption{Acquisition setups used for acquiring EMG signals from study participants. (a) shows the Olimex EMG shield with six channels; (b) shows an armband similar to the Myo armband with eight channels.}
    \label{fig:acquisition_setup}
\end{figure}
Since EMG signals are also commonly recorded using a band form factor, we recorded the third dataset with an EMG band~\footnote{\url{http://www.oymotion.com/en/product32/149}} that has 8 EMG channels and a sampling rate of 500 Hz. The band also provides additional data, such as accelerometer and gyroscope values, but these were not used in the analysis and were, therefore, discarded.

We recorded \rev{the first two datasets} using bipolar wet electrodes of 10 mm size (Covidien, H124SG), which are most commonly used for EMG signal acquisition.

\subsection{Application Scenarios}

\begin{figure*}
    \centering
    \includegraphics[width=1\linewidth]{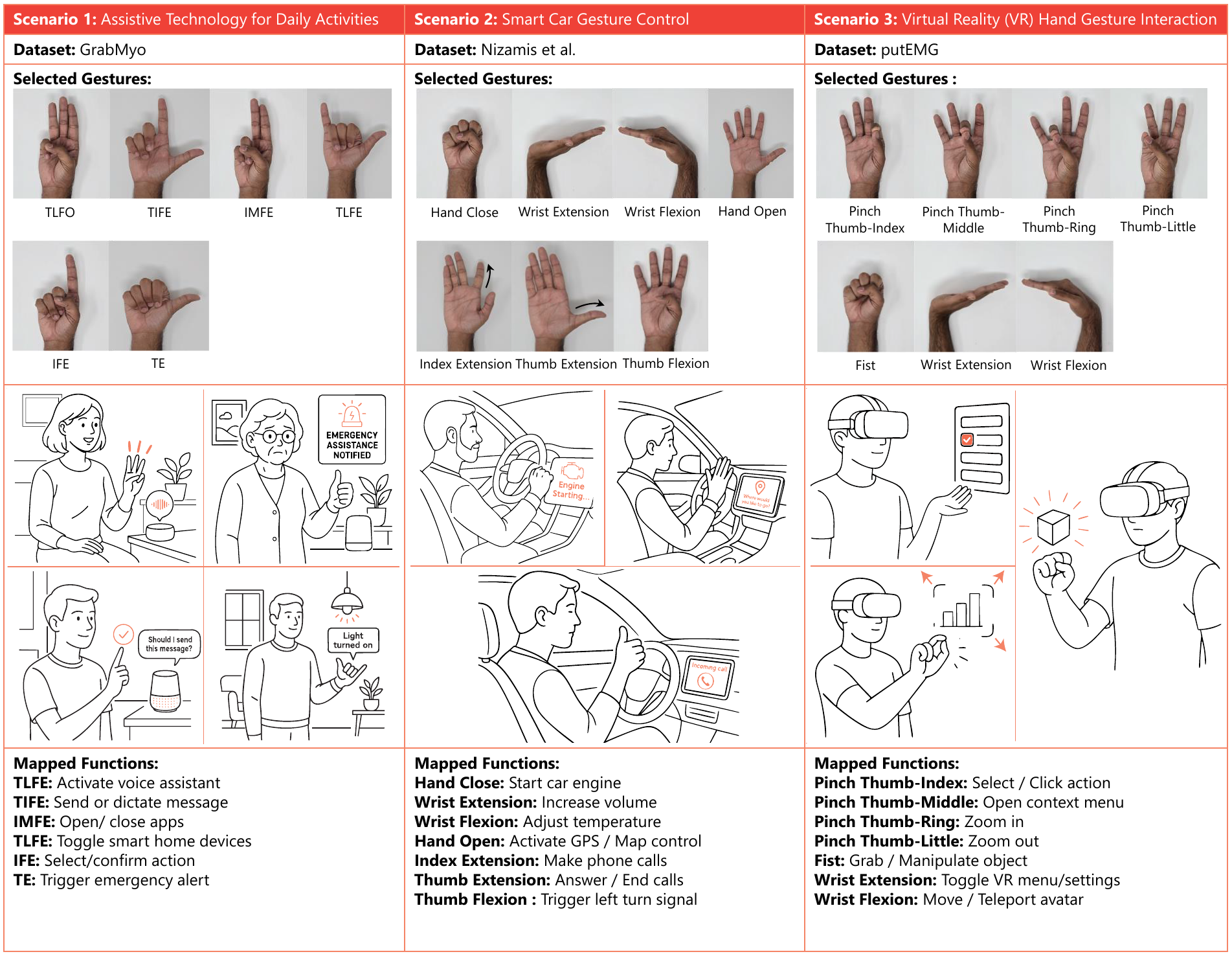}
    \caption{Overview of the three application scenarios used in the study: (1) Assistive Technology for Daily Activities using the GrabMyo \cite{grabmyo} dataset, (2) Smart Car Gesture Control using the Nizamis et al. \cite{nizamis2020characterization} dataset, and (3) Virtual Reality (VR) Hand Gesture Interaction using the putEMG \cite{putemg} dataset. Each scenario highlights selected hand and wrist gestures along with their respective mapped functions, demonstrating diverse use-case applications of gesture recognition systems.}
    \label{fig:scenarios}
\end{figure*}

\subsubsection{Scenario 1: Assistive Technology for Daily Tasks}
We envision our computational design tool assisting a designer, Ben, in creating a gesture recognition system for accessibility. Ben needs to design finger-based gestures that allow elderly users to perform daily tasks intuitively. He selects six gestures from the GrabMyo dataset~\cite{grabmyo}, which are mapped to actions such as controlling smart home devices, initiating calls, sending messages, opening/closing apps, selecting actions, and triggering emergency alerts. Figure~\ref{fig:scenarios} illustrates the gestures and their corresponding functions.

%In this scenario, the hand gestures are mapped to assistive functions designed to help individuals with disabilities perform daily tasks more independently. We selected six gestures from the GrabMyo dataset \cite{grabmyo}. These gestures are mapped to specific assistive actions such as controlling smart home devices, initiating calls, or triggering emergency alerts. 

\subsubsection{Scenario 2: Smart Car Control}

Similarly, another scenario involves Aisha, a designer focused on improving in-vehicle interaction through gesture-based controls. Her goal is to reduce driver distraction by enabling hands-free operation of essential car functions. Using the dataset from Nizamis et al.\cite{nizamis2020characterization}, she selects seven expressive hand and wrist gestures—such as wrist flexion, thumb extension, and fist—to represent commands like starting the engine, adjusting climate or audio settings, and answering phone calls. These gestures are integrated into the vehicle's interface to support intuitive, safe, and responsive interaction. Figure \ref{fig:scenarios} depicts each gesture and its corresponding function.

\subsubsection{Scenario 3: Virtual Reality Interaction}

In this scenario, Ravi, a designer working on immersive interaction systems, explores the use of hand gestures to navigate and interact within a Virtual Reality (VR) environment. His objective is to enhance the intuitiveness of VR experiences by replacing traditional input methods with natural gesture-based controls. Drawing from the putEMG dataset~\cite{putemg}, Ravi selects seven gestures that are well-suited for tasks such as object manipulation, spatial navigation, and engaging with virtual menus. These gestures are seamlessly integrated into the VR interface to support a more immersive and fluid user experience. Figure~\ref{fig:scenarios} illustrates the gesture-action mappings.

\subsection{Participants}

We recruited 8 participants for each scenario, with some participants overlapping between scenarios. The first scenario had 8 participants (4 M, 4 F, mean age: 25.5; SD: 5.1) with an average forearm length of 26.71 cm (SD: 2.02 cm) and an average forearm circumference of 24.57 cm (SD: 3.56 cm). The second scenario had 8 participants (5 M, 3 F, mean age: 26.13; SD: 5.46) with an average forearm length of 27.1 cm (SD: 2.47 cm) and an average forearm circumference of 25.47 cm (SD: 3.5 cm). The third scenario also had 8 participants (4 M, 4 F, mean age: 26; SD: 5.42), with an average forearm length of 28.9 cm (SD: 1.24 cm) and an average forearm circumference of 26.28 cm (SD: 3.5 cm).

\subsection{Procedure}

% While most datasets mentioned in the Table \ref{tab:datasets} use mono-polar electrodes for acquisition, we opted for bi-polar electrodes, as they are more commonly used. As a result, an extra electrode was placed on top of the one selected by the tool in the stencil.

% Stencil generation and placing on arm
We first collected participant-specific measurements (forearm length and circumference at various points as required by each scenario) to generate a customized stencil based on the electrode layout suggested by the tool. This stencil was then sent to a paper cutter to produce a precise paper cutout. The placement of the stencil on the forearm was determined by the guidelines mentioned in the original dataset of the respective scenario. Once aligned according to the measurements, the stencil was carefully applied to the participant’s forearm. Electrode positions were then marked directly onto the skin through the stencil using a highlighter, after which the stencil was removed. After removing the stencil, the electrodes were placed on the user's skin.

% dataset collection
Initially, the participant performed a practice block with each gesture to help them become familiar with the gestures. Each block has one trial per gesture, and 10 blocks were recorded for each participant. After performing each gesture, the participants were asked to bring their hands back to the rest state. This time between two gestures is recorded as a rest (non-gesture) gesture for the classifier. The order of the gestures was randomized within each block. After each block, participants were allowed to take a break and continue when ready. The full data collection session took approximately one hour per participant for each dataset.

Overall, our dataset comprises a total of 2,080 trials $(10\ \text{trials} \times 8\  \text{participants} \times 25\  \text{gestures})$, with gestures drawn from three different datasets as well as a rest state.

\subsection{Feature Extraction and Classification Comparison}

In order to do a systematic comparison, we first evaluated the classification accuracy of the selected gestures using data from the original datasets \rev{(Table \ref{tab:datasets})}. For each dataset, we selected the top electrodes based on the rankings generated by the Permutation Importance + Random Forest (PI + RF) combination (six electrodes for \rev{scenario one and two}, and eight electrodes for \rev{scenario three}) \rev{using data from a random user in the dataset}. \rev{This layout of top electrodes was then used for comparison between the original datasets and the datasets we recorded in our study.}

\rev{For the original dataset,} we used 10 trials per gesture from a single session and computed the classification accuracy using 4-fold cross-validation with a 75\% training and 25\% testing split \rev{for all the users in the dataset}. The same feature vector described in Section~\ref{classifiers} was used as input to the classifier.

\rev{We then evaluated classification accuracy using the datasets recorded in our study, following the same 4-fold cross-validation strategy and using the same feature vector to ensure consistency and fair comparison}. For \rev{each} recorded dataset, we computed the average accuracy across all 8 participants and \rev{compared them with the average accuracies calculated using the original datasets across all users}.

\subsection{Results}

\begin{figure}
    \centering
    \includegraphics[trim={1cm 1cm 0.9cm 1cm},clip,width=1\linewidth]{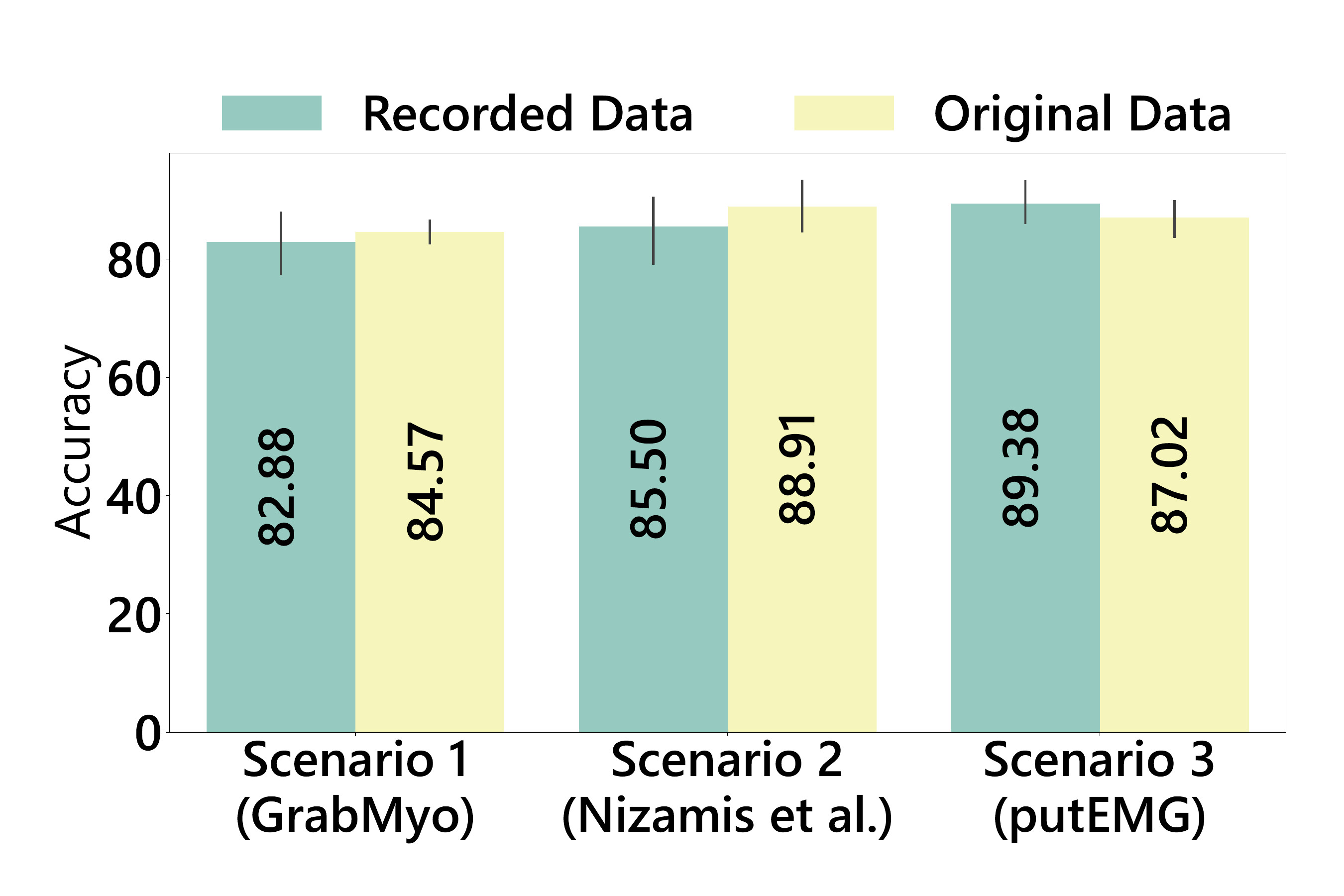}
    \caption{Comparison of accuracy between the recorded data and the data from the original datasets.}
    \label{fig:study_results}
\end{figure}

Figure~\ref{fig:study_results} shows the classification accuracies obtained from the original datasets and the recorded datasets using the selected gesture subsets and electrode layouts generated by the tool.

For \rev{Scenario} 1, the \rev{average} accuracy with the original dataset using the top six electrodes was \rev{84.57\% (SD: 0.07)}, while the recorded dataset achieved an average accuracy of \rev{82.88\% (SD: 0.08)}. For \rev{Scenario} 2, accuracy was \rev{88.91\% (SD: 0.06)} for the original dataset, compared to \rev{85.5\% (SD: 0.08)} for the recorded dataset. In \rev{Scenario} 3, which used eight electrodes, accuracy was \rev{87.02\% (SD: 0.11)} for the original data and \rev{89.38\% (SD: 0.05)} for the recorded data \rev{which was higher than the original dataset.}

Overall, the recorded datasets achieved accuracy levels comparable to the original datasets, with a drop of 2-4\%. This indicates that the electrode layouts generated by the SparseEMG design tool generalized well to real-world hardware and participant variations. These results demonstrate the practical applicability and reliability of the tool for sparse EMG layout generation in real-world settings.

%% file: discussion.tex
\section{Discussion, Limitations and Future Work}
SparseEMG introduces a novel approach to EMG-based interaction by optimizing electrode layouts for sparsity without significantly compromising gesture recognition accuracy. A key contribution is demonstrating that the combination of Permutation Importance (PI) and Random Forest (RF) provides a strong trade-off between these factors across diverse datasets, outperforming other techniques by maintaining a high classification accuracy (94.04\%) while reducing the number of electrodes by 53.5\%. This balance is critical for real-world applications where hardware limitations and user comfort are paramount. Unlike prior EMG toolkits, SparseEMG emphasizes sparse electrode configurations in achieving practical, scalable myoelectric systems, abstracting the complexities of electrode selection to empower designers to focus on gesture creation rather than underlying technical challenges, which could accelerate the adoption of myoelectric interfaces in fields such as assistive technologies, gaming, and augmented reality.

\rev{
\paragraph{\textbf{Benefits and Practical Usability}}: SparseEMG provides major benefits by optimizing the electrode count as follows:}
\begin{itemize}[leftmargin=*]
    \item \rev{Cost and Electrode Reduction: Many low-cost EMG systems (e.g. Arduino-based) support up to 6 channels for $\sim$\$100. Scaling to 8 channels raises costs to $\sim$\$500–$\$$1.5K, while 16- and 64-channel lab-grade setups start at $\sim$\$8K and $\sim$\$10K–\$25K respectively (excluding amplifiers/electrodes) and require specialized training. Stacking Arduinos is possible but impractical due to time sync and frame rate issues. Reducing channel count helps avoid these challenges. Since each channel uses 3 electrodes, even reducing from 8 to 6 channels removes 6 electrodes. As shown in Figure \ref{fig:20_electrode_combination} via Sparsity Score, our tool can reduce total channels from 192 to 11 with only $\sim$7-8\% drop in accuracy.}

    \item \rev{Boosting Frame Rate: For Arduino-based setups, having fewer channels means higher frame rate. With a standard 10-bit ADC the Arduino takes $\sim$100-110 ms to sample each channel. Hence, if the number of channels is reduced from 6 to 4, the frame rate can be boosted from $\sim$240 fps to $\sim$360 fps (assuming a baud rate of 115200).}

    \item \rev{Simplified \& Scalable Electrode Placement: Electrode placement is critical for EMG signal acquisition but often requires expertise in anatomy, biomedical engineering, and consideration of the gesture set and ML model. Our tool abstracts these complexities, enabling users to generate gesture-specific layouts with just a few clicks. Scaling across users is difficult due to anatomical differences, so prior work relied on one-size-fits-all solutions like arrays or bands. Our stencil-generation feature enables reproducible and user-specific placements, which, to our knowledge, has not been explored in prior EMG classification work.}
\end{itemize}

\paragraph{\textbf{Scalablility Across Users and Hardware Platforms}}: The generalizability of SparseEMG's recommendations across datasets and users further highlights its robustness. Our evaluation reveals that sparse layouts derived from one user or dataset perform consistently well when applied to others, with accuracy drops typically within 2–4\%. This suggests that the tool generates electrode configurations that are not overly dependent on individual-specific or dataset-specific variations. This scalability across hardware platforms, from high-density arrays to consumer-grade devices, showcases its adaptability for both research and practical applications, ensuring its relevance across academic studies to commercial products. Moreover, SparseEMG positions machine learning as a design material, enabling interaction designers to iteratively prototype gesture sets and electrode configurations. By providing real-time feedback on estimated gesture recognition accuracy, the tool supports the rapid exploration of design alternatives, such as refining gesture sets or optimizing electrode placement for ergonomic and functional constraints, aligning with the broader HCI goal of leveraging computational tools to augment the creative process.

\paragraph{\textbf{Improving Cross-User Performance}}: While SparseEMG offers significant advancements, several limitations warrant discussion. First, our focus was on user-dependent models, and we did not evaluate whether the selected combination performs effectively in user-independent scenarios due to the limited number of users in the datasets. Building a robust user-independent model would require more diverse data, including samples from children, elderly individuals, and other underrepresented groups. Future work could explore integrating user-independent classifiers or transfer learning techniques to improve cross-user performance. Second, the tool currently supports gesture selection within a single dataset, which restricts its ability to handle gestures aggregated from multiple datasets. This limitation stems from variations in electrode placements, hardware configurations, and recording protocols across datasets. A standardized framework for integrating heterogeneous datasets could address this challenge and further enhance the tool’s versatility. Third, while the tool abstracts the complexities of electrode placement, it relies on user-provided measurements for stencil generation. Variations in forearm dimensions and electrode placement guidelines across datasets introduce potential inconsistencies that may impact performance. Incorporating automated calibration methods or adaptive placement strategies could mitigate these issues.

\paragraph{\textbf{Integrating Deep Learning Architectures}}: SparseEMG opens several exciting avenues for future research. First, incorporating deep learning architectures, such as convolutional or recurrent neural networks, could enhance the tool’s ability to model complex spatiotemporal dependencies in EMG signals. While our current approach prioritizes computational efficiency, advancements in hardware capabilities may enable the integration of more sophisticated models without compromising real-time performance. Second, expanding the tool’s functionality to support dataset-independent modeling would significantly broaden its applicability. This could involve developing meta-learning algorithms that generalize across datasets or incorporating domain adaptation techniques to handle variations in hardware and recording conditions. 

\rev{\paragraph{\textbf{Applications beyond EMG}}:This work focuses on EMG gesture recognition, but the approach can be applied more broadly within the general category of Muscle-Computer Interfaces as introduced by Saponas et al.~\cite{saponas_muscle_computer_chi10}. Additionally, it can be adapted for use with various other sensing technologies, such as inertial measurement units (IMUs), electrical sensing, and mm-wave radars. For example, while our primary focus is on gesture recognition, the placement of electrodes is also crucial for providing haptic feedback through electrical muscle stimulation (EMS).

Moreover, the design tool can be adjusted for other applications, including muscle fatigue assessment, rehabilitation, movement analysis, and prosthesis control. Finally, the SparseEMG framework can be extended to incorporate other biosignal modalities, such as electroencephalography (EEG) and IMU sensors. By generalizing the design principles of our tool, we envision a wider array of computational tools that will empower designers to create interactive systems across multiple sensing technologies.

}
% \subsection{Gesture Selection Across Datasets}
% The tool only allows selection of gestures from one dataset and not across-datasets. This is because the datasets have different recording setups

% \subsection{Deep Learning Architectures}
% Although we tested one MLP, we didn't test how well the more complex deep learning architectures such as LSTM, CNNs etc perform as the main the aim of the tool to make the tool rapid 

% \subsection{Placement specific to User Body Dimensions}
% Each paper has different recording setups and not all papers mention the exact method of placing their recording setups on the forearm, for eg \cite{hyser} only says that they positioned the arrays at the center of the forearm of the user, but didn't give any dimensions of the forearm which makes it difficult to replicate the layout for other users

% \subsection{User-Independent Models}
% We are only doing user dependent models right now

% \subsection{Dataset Independent Modelling}

% \subsection{Variations in Datasets}
% electrode placements, hardware, users, applications (HCI vs prosthesis)

% \subsection{Adding Additional Datasets}

\section{Conclusion}

% In this work, we systematically explored the relationship between electrode selection strategies and machine learning classifiers in the context of EMG-based gesture recognition. By evaluating 28 combinations, comprising four electrode selection schemes and seven classifiers, across six diverse datasets, we demonstrated that electrode count can be reduced by up to 53\% without significant drop in classification accuracy. Our results highlight the effectiveness of using the Permutation Importance method for electrode selection paired with a Random Forest classifier. Building on these insights, we introduced SparseEMG, a novel design tool that enables users to generate sparse electrode layouts tailored to specific gesture sets, hardware constraints, and machine learning configurations. SparseEMG supports 50+ unique gestures and our validation experiments confirm that (1) electrode layouts suggested by our tool are scalable across users and (2) the tool’s practical usefulness and generalizability to three real-world applications
% This work underscores the significance of informed electrode selection and contributes a scalable, data-driven tool for developing efficient and accurate EMG-based gesture recognition systems. SparseEMG fills a crucial and critical gap by focusing on rapid and interactive gesture design for EMG applications. We believe this will enable more widespread adoption of EMG for practical gesture recognition.

In this work, we systematically explored the relationship between electrode selection strategies and machine learning classifiers in the context of EMG-based gesture recognition. By evaluating 28 combinations, comprising four electrode selection schemes and seven classifiers, across six diverse datasets, we demonstrated that electrode count can be reduced by up to 53\% without a significant drop in classification accuracy. Our results highlight the effectiveness of using the Permutation Importance method for electrode selection, paired with a Random Forest classifier. Building on these insights, we introduced SparseEMG, a novel design tool that enables users to generate sparse electrode layouts tailored to specific gesture sets, hardware constraints, and machine learning configurations. SparseEMG supports over 50 unique gestures, and our validation experiments confirm that: (1) the electrode layouts generated by the tool scale effectively across users, and (2) the tool demonstrates strong practical utility and generalizability across three real-world application scenarios. This work advances the field by foregrounding the value of principled electrode selection and contributes a data-driven, user-centered tool to accelerate the development of efficient EMG-based interaction systems. By enabling rapid iteration and informed trade-offs, SparseEMG addresses a critical bottleneck in EMG interface design and lays the groundwork for broader deployment of EMG in everyday interactive technologies.